\documentclass{iopjournal}
\usepackage{graphicx}
\usepackage{amsfonts}
\usepackage{amsmath}
\usepackage{float}
\usepackage{subcaption}
\usepackage{multicol}
\usepackage{ragged2e}
\usepackage{xcolor}
\usepackage{hyperref}
\usepackage[numbers,sort&compress]{natbib}

\begin{document}
\justifying

\articletype{paper} %	 e.g. Paper, Letter, Topical Review...

\title{Self-consistent analysis of the Kuramoto model with higher-order interactions}

\author{Chanin Kumpeerakij$^{1,2}$, Juan G. Restrepo$^3$}

\affil{$^1$Department of Physics, University of Colorado, Boulder, Colorado, USA} \\
\affil{$^2$Interdisciplinary Quantitative Biology Program (IQ Biology), BioFrontiers Institute, University of Colorado, Boulder, Colorado, USA}\\
\affil{$^3$Department of Applied Mathematics, University of Colorado at Boulder, CO, USA}

\email{$^1$chanin.kumpeerakij@colorado.edu, $^2$juanga@colorado.edu}


\begin{abstract}
\justifying
The Kuramoto model with higher-order interactions has recently been shown to exhibit bistability, explosive synchronization transitions, and rich collective dynamics. Existing analytical approaches, however, typically rely on all-to-all coupling or mean-field approximations of the underlying hypergraph structure. While these methods describe typical networks in the thermodynamic limit, they generally fail to capture the effects of finite hypergraph and oscillator frequency realizations. To address this limitation, we develop a self-consistent analytical framework for the Kuramoto model with dyadic and  triadic interactions on hypergraphs. We introduce generalized local order parameters that capture the combined effects of dyadic and triadic phase correlations, and derive a hierarchy of approximation schemes for the local and global synchronization order parameters. Using these approximations, we determine critical coupling strengths for the onset of synchronization and bistability. In particular, we show that the critical triadic coupling strength governing the onset of bistability depends on correlations between the eigenvectors of the dyadic adjacency matrix and the triadic interaction structure. Numerical simulations on homogeneous and heterogeneous hypergraphs validate the theory and illustrate the distinct regimes of applicability of the approximation schemes.
\end{abstract}
\section{Introduction} \label{sec:introduction} \

The study of synchronization in systems of coupled oscillators has long served as a cornerstone of nonlinear dynamics, with applications ranging from power grid stability \cite{Rohden2012} and neurobiological rhythms \cite{Varela2001, Breakspear2017} to social coordination \cite{Neda2000, Castellano2009}. For decades, the Kuramoto model for synchronization of phase oscillators has provided a robust and flexible framework for understanding these phenomena through pairwise (dyadic) interactions \cite{KMreview, Kuramoto, Strogatz2000}. However, dyadic models cannot represent interactions in which groups of three or more oscillators interact jointly and non-additively (higher-order interactions), which are present in many real-world systems, such as the synchronized firing of neuronal groups or multi-agent consensus in social networks \cite{Iacopini2024, Yu2011, Combrisson2025, Bullmore2009}. Higher-order interactions also arise naturally in phase reduction procedures when weak coupling expansions are carried beyond the first order \cite{ASHWIN201614,Leon2019}. These developments  have motivated the study of synchronization mediated by higher-order interactions, typically modeled using hypergraphs or simplicial complexes \cite{Battiston2020, Skardal2020, Skardal2019, Skardal2021, Xu2020, Sabhahit2024}. Higher-order interactions in coupled oscillator models are known to induce novel phenomena like explosive synchronization and bistability \cite{Skardal2019, Adhikari2023} (similar phenomena are also observed in social contagion \cite{Iacopini2019,DeArruda2020, Landry2022} and other systems \cite{Grilli2017}). 

While the effect of higher-order interactions on the synchronization transition is relatively well understood for all-to-all coupled systems or systems that can be described with mean-field approximations~\cite{Skardal2020, Wang2021, XU2023113343, Sabhahit2024, Adhikari2023}, there is currently no self-consistent analytical framework for higher-order Kuramoto models that systematically incorporates finite hypergraph and oscillator frequency realizations. In this paper we develop such a framework. Below we introduce the Kuramoto model with higher-order interactions that will be studied in this paper, and discuss some of the previous work.

\subsection{Kuramoto Model with higher-order interactions}  \

We consider the Kuramoto model on a network with higher-order interactions, where the phase $\theta_n$ of oscillator $n$ evolves as 

\begin{align}
    \frac{d\theta_n}{dt} = \omega_n + K_2 \sum^N_{m=1}\boldsymbol{A}_{nm}\sin(\theta_m-\theta_n)  + K_3 \sum^N_{m=1}\sum^N_{j=1}\boldsymbol{B}_{nmj} \sin(2\theta_m-\theta_j-\theta_n),  \label{eq:ode_sys}
\end{align}
where $n = 1,2,...,N$, $\omega_n$ is the intrinsic frequency of oscillator $n$, and $K_2$ and $K_3$ are the coupling strengths for dyadic and triadic interactions, respectively. The adjacency matrix $\boldsymbol{A}_{nm}$ and tensor $\boldsymbol{B}_{nmj}$ encode the structures of the dyadic and triadic interactions, respectively: $\boldsymbol{A}_{nm}>0$ if oscillator $m$ is coupled to oscillator $n$ via a dyadic interaction, and $\boldsymbol{A}_{nm}=0$ otherwise; similarly, $\boldsymbol{B}_{nmj}>0$ if oscillators $m$ and $j$ are coupled to oscillator $n$ via a triadic interaction, and $\boldsymbol{B}_{nmj}=0$ otherwise.

We note that the specific form of triadic coupling in Eq. (\ref{eq:ode_sys}) is one of the types that arise when the phase reduction procedure that usually leads to the Kuramoto model with dyadic interactions is carried beyond the first order in the coupling strength \cite{Leon2019}. While we chose only one such type of coupling for simplicity, we expect that our method could be adapted to other coupling types. We emphasize, however, that all of the results below, in particular the expressions for the triadic order parameter $R^{(3)}_n$ and for the critical triadic coupling strength $K_3^c$, are specific to the coupling in Eq.~(\ref{eq:ode_sys}) without phase lags. If a phase lag is included in the coupling functions, as in the Kuramoto--Sakaguchi model, the drifting oscillators also contribute to the dyadic order parameter and the analysis below must be modified. Similarly, other triadic couplings, such as $\sin(\theta_m+\theta_j-2\theta_n)$, which have been studied in Refs.~\cite{Carballosa2023, Leon2024, Marui2025}, lead to different self-consistent equations and possibly to different conclusions regarding bistability. Finally, while we take the interaction weights to be non-negative, the TAT and FDA equations derived below only involve sums over the couplings and hold formally for weights of either sign; positivity is used in the Perron--Frobenius argument of Sec.~\ref{sec:PT_BI} and in the single-cluster assumption, which may fail for signed couplings that favor anti-phase configurations.

\subsection{Previous Work} \

Various versions of Eq.~(\ref{eq:ode_sys}) have been studied extensively, so we summarize only the main relevant results. The original Kuramoto model, for which $\boldsymbol{B}_{nmj} = 0$, $\boldsymbol{A}_{nm} = 1$, and $K_2 = 1/N$, was originally studied by Kuramoto using a self-consistent analysis  \cite{KMreview, Kuramoto, Strogatz2000}, and solved in the $N\to \infty$ limit by Ott and Antonsen \cite{Ott2008}. The Kuramoto model on dyadic networks, for which $\boldsymbol{B}_{nmj} = 0$, has been studied using self-consistent methods \cite{Restrepo2005}, heterogeneous mean-field approximations \cite{Restrepo2005,Ichinomiya2004,Lee2005}, and by adapting the all-to-all Ott-Antonsen Ansatz to the network case \cite{Barlev2011,Restrepo_2014}. Ref.~\cite{RODRIGUES20161} provides a review of work on the Kuramoto model on dyadic networks.

More recently, the Kuramoto model with higher-order interactions [Eq.~(\ref{eq:ode_sys}) with $\boldsymbol{B} \neq 0$] has been solved exactly using the Ott-Antonsen Ansatz in the all-to-all case, $\boldsymbol{B}_{nmj} = 1$, $\boldsymbol{A}_{nm} = 1$, by Skardal and Arenas \cite{Skardal2020}. An alternative method of solution for the all-to-all case that uses a self-consistent approach has recently been proposed in Refs.~\cite{Wang2021,XU2023113343,Sabhahit2024}. 
In order to study networks that are not fully connected, Ref.~\cite{Adhikari2023} combines the Ott-Antonsen Ansatz with a Heterogeneous Mean-Field approximation. This technique allows for the study of the effect of heterogeneous degree distributions and correlations between dyadic and triadic interaction structures on the synchronization transition. 
%However, due to its mean-field nature, this method does not capture the effect of specific hypergraph and frequency realizations, and is only valid in the $N \to \infty$ limit.
The Kuramoto model with higher-order interactions has also been studied in simplicial complexes \cite{Skardal2019}, in particular in the case where phase signals are assigned not only to the nodes of the network but also to faces of the simplicial complex \cite{Millan2020}.

Despite these previous results, the effect of specific hypergraph and frequency realizations on the synchronization transition is not well understood. In this work we propose a self-consistent framework extending the approach developed by Ref.~\cite{Restrepo2005} and Refs.~\cite{Sabhahit2024,Wang2021,XU2023113343} to describe the synchronization dynamics of oscillators on hypergraphs. 

The remainder of this paper is organized as follows. In Section \ref{sec:self-consistency}, we develop a self-consistent method to estimate the system's order parameters and a sequence of approximations that are successively less accurate but require less information about the system. In Section \ref{sec:PT_BI} we use this analytical framework to find the values of the dyadic and triadic coupling strengths determining the onsets of synchronization and bistability, respectively. We explicitly summarize the underlying assumptions of these methods in Section \ref{assumptions}. In Section \ref{sec:validation}, we validate our results using numerical simulations on hypergraphs with both uniform and power-law degree distributions. We specifically examine the onset of bistability and explosive synchronization. Finally, Section \ref{sec:discussion} provides a discussion of our results and concluding remarks. 

\newpage
\section{Self-consistent Analysis}\label{sec:self-consistency} \

We start our analysis by defining the local dyadic and triadic order parameters,
\begin{align}
     R^{(2)}_ne^{i\psi^{(2)}_n} &= \sum^N_{m=1}\boldsymbol{A}_{nm}e^{i\theta_m}, \label{eq:local_R2_def}\\ 
    R^{(3)}_ne^{i\psi^{(3)}_n} &= \sum^N_{m=1}\sum^N_{j=1} \boldsymbol{B}_{nmj}e^{i(2\theta_m-\theta_j)}.\label{eq:local_R3_def}
\end{align}

With Eqs. (\ref{eq:local_R2_def}) and (\ref{eq:local_R3_def}), we can reformulate Eq. (\ref{eq:ode_sys}) as:
\begin{align}
    \frac{d\theta_n}{dt} &= \omega_n- H_n\sin(\theta_n - \psi_n), \label{eq:reduce_dyn_eq}
\end{align}
where we defined
\begin{align}\label{eq:hn}
	H_ne^{i\psi_n} &= K_2R^{(2)}_ne^{i\psi_n^{(2)}} + K_3 R^{(3)}_ne^{i\psi^{(3)}_n}
\end{align} 
(see Appendix~\ref{appx:reduce_dyn} for details).

In the following subsections, we will develop a hierarchy of approximations to determine the order parameters given $K_2$ and $K_3$. These approximations will become less accurate as the amount of information about the system that they incorporate decreases. In order to develop these approximations, we will make a series of assumptions that will be explicitly listed and discussed in Sec.~\ref{assumptions}.

\subsection{Time-Averaged Theory} \

Equation (\ref{eq:reduce_dyn_eq}) can be treated in the same way in which Kuramoto treated the all-to-all dyadic case~\cite{KMreview, Kuramoto,Strogatz2000}: we seek a solution in which both $H_n$ and $\psi_n$ are constant. In reality, for finite networks, $H_n$ and $\psi_n$ fluctuate over time, and therefore we effectively assume that the time fluctuations of these quantities are negligible. Following the convention in Ref.~\cite{Restrepo2005}, we refer to this approximation as the {\it Time-Averaged Theory} (TAT). In this approximation, the phase angle $\theta_n$ governed by Eq.~(\ref{eq:reduce_dyn_eq}) has two possible long-term behaviors: if $|\omega_n| \leq H_n$, then $\theta_n$ will asymptotically approach the stable fixed point $\theta^*_n$ satisfying
\begin{align}
\sin(\theta^*_n - \psi_n) = \frac{\omega_n}{H_n},
\end{align}
in which case oscillator $n$ is denoted as a {\it locked oscillator}. If $|\omega_n| > H_n$, then $\theta_n - \psi_n$ will always increase or decrease, depending on the sign of $\omega_n$, and oscillator $n$ is said to be {\it drifting}.

In addition to assuming that the phase angles $\psi_n$ are constant, we will introduce the additional assumption that they are all equal. Since the system is rotationally invariant, the constant can be chosen to be zero, $\psi_n = 0$, so that the locked oscillators satisfy $\sin(\theta^*_n) = \omega_n/H_n$. This is the form used in the remainder of the paper. The assumption that all the angles $\psi_n$ are equal amounts to postulating that all the oscillators form a single synchronization cluster. This assumption will not necessarily be valid in networks with community or spatial structure, where different communities or spatial locations can give rise to different synchronization clusters. However, we expect that our analysis method could be extended to some of these situations.

% The analysis indicates that oscillators are in a synchronized phase and a coherent state when $|\omega_n| \leq H_n$. Conversely, oscillators are in a drifting phase, undergoing random phase evolution, when $|\omega_n| > H_n$. Subsequently, we divide each term in $H_n$ into distinct categories based on the type of oscillator, as previously described.

% We now apply the similar process as in Eq. \ref{eq:r_split} with the assumption that the system achieves long-term stability, $H_n(t)$ remains constant, and $\psi_n(t)$ rotates at a uniform frequency $\Omega$. Consequently, the system can be transformed into a rotating frame, allowing $\psi_n$ to be set to zero without loss of generality. This transformation also results in $\psi^{(2)}_n = \psi^{(3)}_n = 0$. Subsequently, the summation terms of $R^{(2)}_n$ and $R^{(3)}_n$ are considered to represent contributions from locked and drifted phase oscillators.

Now we separate the oscillators in the sums in Eqs.~(\ref{eq:local_R2_def}) and (\ref{eq:local_R3_def}) into locked and drifting oscillators. 
\begin{align}
    R^{(2)}_n &= \sum_{|\omega_m| \le H_m }\boldsymbol{A}_{nm}e^{i\theta_m} +\sum_{|\omega_m| > H_m}\boldsymbol{A}_{nm}e^{i\theta_m}, \label{eq:r2_sum}\\
    R^{(3)}_n &= \sum_{\substack{|\omega_m| \le H_m \\ |\omega_j| \le H_j}}\boldsymbol{B}_{nmj}e^{i(2\theta_m-\theta_j)} + \sum_{\substack{|\omega_m| > H_m \\ |\omega_j| \le H_j}}\boldsymbol{B}_{nmj}e^{i(2\theta_m-\theta_j)}  \nonumber\\
    &+\sum_{\substack{|\omega_m| \le H_m \\ |\omega_j| > H_j}}\boldsymbol{B}_{nmj}e^{i(2\theta_m-\theta_j)}+\sum_{\substack{|\omega_m| > H_m \\ |\omega_j| > H_j}}\boldsymbol{B}_{nmj}e^{i(2\theta_m-\theta_j)}.\label{eq:r3_sum}
\end{align}
Next, we take a long-time average, indicated by $\langle \cdot \rangle_t$. For the dyadic order parameter, $R^{(2)}_n$, we obtain
\begin{align}
	\langle R^{(2)}_n \rangle_t&= \sum_{|\omega_m|\le H_m} \boldsymbol{A}_{nm} \langle e^{i\theta_m}\rangle_t + \sum_{|\omega_m| > H_m} \boldsymbol{A}_{nm} \langle e^{i\theta_m}\rangle_t \\
    &= \sum_{|\omega_m|\le H_m} \boldsymbol{A}_{nm} \langle \cos(\theta_m)\rangle_t +i\sum_{|\omega_m| \le H_m} \boldsymbol{A}_{nm} \langle \sin(\theta_m)\rangle_t + \sum_{|\omega_m| > H_m} \boldsymbol{A}_{nm} \langle e^{i\theta_m}\rangle_t, \\
        &= \sum_{|\omega_m|\le H_m} \boldsymbol{A}_{nm}  \cos(\theta^*_m) +i\sum_{|\omega_m| \le H_m} \boldsymbol{A}_{nm} \sin(\theta^*_m) + \sum_{|\omega_m| > H_m} \boldsymbol{A}_{nm} \langle e^{i\theta_m}\rangle_t, \\
                &= \sum_{|\omega_m|\le H_m} \boldsymbol{A}_{nm}\sqrt{1-\left(\frac{\omega_m}{H_m} \right)^2}  +i\sum_{|\omega_m| \le H_m} \boldsymbol{A}_{nm} \left(\frac{\omega_m}{H_m}\right) + \sum_{|\omega_m| > H_m} \boldsymbol{A}_{nm} \langle e^{i\theta_m}\rangle_t. \label{eq:each_term_R2}
\end{align}

To make further progress, we assume that the natural frequency $\omega_n$ of node $n$ and its local order parameter $H_n$ are statistically independent. Note that, while $H_n$ does not depend explicitly on $\omega_n$, it depends on the phases of the neighbors of node $n$.
In dense networks, these neighbors are typically connected to many other oscillators, so correlations between their phases and $\omega_n$ are expected to be weak. 
However, in highly heterogeneous networks, particularly near the onset of synchronization, highly connected nodes may disproportionately influence their neighbors. This can induce correlations between $\omega_n$ and $H_n$. 
Therefore, we expect this approximation to be most accurate for dense networks that are not too heterogeneous. The validity of the approximation that $H_n$ is statistically independent from $\omega_n$ is tested numerically in Appendix \ref{sec:real-term-appx}.

Using this approximation, if the frequency distribution is assumed to be symmetric, the second and third terms can be shown to vanish in the limit when the number of terms in the sum is large (see Appendix \ref{sec:real-term-appx}). Under these assumptions, we obtain
\begin{align}
	R^{(2)}_{n}=  \sum_{|\omega_m|\le H_m} \boldsymbol{A}_{nm} \sqrt{1-\left(\frac{\omega_m}{H_m} \right)^2}\label{eq:r2_tat},
\end{align}
where, for simplicity of notation, we have dropped the $\langle\cdot \rangle_t$ from $R^{(2)}_n$.

Eq.~(\ref{eq:r2_tat}) expresses the dyadic order parameter $R^{(2)}_n$ as a function of the dyadic network structure, the specific collection of intrinsic frequencies, and the quantity $H_n = K_2 R^{(2)}_n + K_3 R^{(3)}_n$, and it is a direct generalization of the analogous result for dyadic interactions in Ref.~\cite{Restrepo2005}. To obtain a closed system of equations, we need to repeat this process for $R^{(3)}_n$. Doing so, we obtain, again dropping the $\langle\cdot \rangle_t$ from $\langle R_n^{(3)}\rangle_t$,
\begin{align}
R_n^{(3)} = R_n^{(3)}(LL) + R_n^{(3)}(LD) + R_n^{(3)}(DL) + R_n^{(3)}(DD),\label{RLDL}
\end{align}
where $LL$, $LD$, $DL$, and $DD$ indicate the contributions where oscillators $m$ and $j$ are both locked, locked and drifting, drifting and locked, and both drifting, respectively. These are given by
\begin{align}
R_n^{(3)}(LL) &= \sum_{\substack{|\omega_m| \le H_m \\ |\omega_j| \le H_j}}\boldsymbol{B}_{nmj}\langle e^{ 2i\theta_m} e^{-i\theta_j}\rangle_t, \label{eq:R3_LL}\\
R_n^{(3)}(LD) &= \sum_{\substack{|\omega_m| \le H_m \\ |\omega_j| > H_j}}\boldsymbol{B}_{nmj}\langle e^{ 2i\theta_m} e^{-i\theta_j}\rangle_t,\label{eq:R3_LD}\\
R_n^{(3)}(DL) &= \sum_{\substack{|\omega_m| > H_m \\ |\omega_j| \le H_j}}\boldsymbol{B}_{nmj}\langle e^{ 2i\theta_m} e^{-i\theta_j}\rangle_t,\label{eq:R3_DL}\\
R_n^{(3)}(DD) &= \sum_{\substack{|\omega_m| > H_m \\ |\omega_j| > H_j}}\boldsymbol{B}_{nmj}\langle e^{ 2i\theta_m} e^{-i\theta_j}\rangle_t. \label{eq:R3_DD}
\end{align}

Now we make the additional approximation that pair correlations can be neglected, as in Ref.~\cite{Adhikari2023}. With this approximation one can replace $\langle e^{ 2i\theta_m} e^{-i\theta_j}\rangle_t$ by $\langle e^{ 2i\theta_m}\rangle_t \langle e^{-i\theta_j}\rangle_t$. For locked oscillators, $\langle e^{ 2 i\theta_j}\rangle_t = e^{ 2 i\theta^*_m}$ and $\langle e^{- i\theta_j}\rangle_t = e^{ - i\theta^*_m}$. For the drifting oscillators we replace the time average with an average over their stationary density (see Appendix~\ref{sec:r3_tat_system} for details). Doing so, we obtain the closed system of equations which constitute the TAT approximation:
\begin{align}
    H_n &= K_2 R^{(2)}_{n} + K_3 R^{(3)}_{n}\label{eq:h_tat_system},\\
	R^{(2)}_{n} &=  \sum_{|\omega_m|\le H_m} \boldsymbol{A}_{nm} \sqrt{1-\left(\frac{\omega_m}{H_m} \right)^2}\label{eq:r2_tat_system},\\
	 R^{(3)}_n &=  \sum_{\substack{|\omega_m| \le H_m \\ |\omega_j| \le H_j}}\boldsymbol{B}_{nmj}\left[1-2\left(\frac{\omega_m}{H_m} \right)^2\right]\sqrt{1-\left(\frac{\omega_j}{H_j} \right)^2} + \label{eq:r3_tat_system}\\\nonumber 
    & \sum_{\substack{|\omega_m| > H_m \\ |\omega_j| \le H_j}}\boldsymbol{B}_{nmj} \left[\frac{2|\omega_m|}{H_m^2}\left(\sqrt{\omega_m^2-H_m^2}-\omega_m\right)+1 \right]\sqrt{1-\left(\frac{\omega_j}{H_j} \right)^2} .
\end{align} \

Remarkably, in contrast to the dyadic Kuramoto model, and as noted in Refs. \cite{Sabhahit2024, Wang2021}, the drifting oscillators do contribute to the order parameter. Having the frequencies of individual oscillators, $\{\omega_n\}^N_{n=1}$, and the adjacency matrix and tensor ${\bf A}$ and ${\bf B}$, one can solve these equations numerically to obtain the local order parameters $R_n^{(2)}$ and $R_n^{(3)}$ as a function of $K_2$ and $K_3$. 

\subsection{Frequency Distribution Approximation} \

The TAT approximation takes into account the frequencies of individual oscillators, and therefore is able to capture the effect of correlations of these frequencies with the structure of dyadic and triadic interactions. In practice, however, one might not have information on individual oscillator frequencies. Therefore, we now average over the frequencies of individual oscillators. We assume that the sums in Eqs.~(\ref{eq:r2_tat_system})-(\ref{eq:r3_tat_system}) contain enough terms to be approximated by expected values over the frequency distribution $g(\omega)$, and that $\omega_m$ and $H_m$ are statistically independent, as introduced above. The sums can then be replaced by integrals as follows:
\begin{align}
\sum_{|\omega_m| \leq H_m} \boldsymbol{A}_{nm} F(\omega_m,H_m) \longrightarrow \sum_{m} \boldsymbol{A}_{nm} \int_{-H_m}^{H_m}  F(\omega,H_m)g(\omega)d\omega,
\end{align}
and, similarly,
\begin{align}
&\sum_{\substack{|\omega_m| \le H_m \\ |\omega_j| \le H_j}} \boldsymbol{B}_{nmj} F(\omega_m,\omega_j,H_m,H_j) \longrightarrow \sum_{m,j} \boldsymbol{B}_{nmj}\int_{-H_m}^{H_m}\int_{-H_j}^{H_j}  F(\omega,\omega',H_m,H_j)g(\omega)g(\omega')d\omega' d\omega,\label{intapp1}\\
&\sum_{\substack{|\omega_m| \ge H_m \\ |\omega_j| \le H_j}} \boldsymbol{B}_{nmj} F(\omega_m,\omega_j,H_m,H_j) \longrightarrow \sum_{m,j} \boldsymbol{B}_{nmj} \int_{|\omega'|> H_m}\int_{-H_j}^{H_j} F(\omega,\omega',H_m,H_j)g(\omega)g(\omega')d\omega' d\omega,\label{intapp2}
\end{align}
where $F$ represents the functions inside the sums in Eqs.~(\ref{eq:r2_tat_system})-(\ref{eq:r3_tat_system}).

For the particular case of a Lorentzian frequency distribution with mean $0$ and width $1$, $g(\omega) = 1/[\pi(1+\omega^2)]$, doing the integrals (see Appendix \ref{appx:FDA_integral}) we obtain the {\it Frequency Distribution Approximation} (FDA), which is the closed system of equations given by 
 \begin{align}
    H_n &= K_2 R^{(2)}_{n} + K_3 R^{(3)}_{n}\label{eq:h_fda_system},\\
 	R^{(2)}_n &= \sum_m \boldsymbol{A}_{nm} \frac{\sqrt{H_m^2+1}-1}{H_m} \label{eq:r2_fda}, \\
     R^{(3)}_n &= \sum_{m,j} \boldsymbol{B}_{nmj} \left(\frac{\sqrt{H_m^2+1}-1}{H_m}\right)^2 \left( \frac{\sqrt{H_j^2+1}-1}{H_j} \right) \label{eq:r3_fda}.
 \end{align}
 
 The FDA approximates the local order parameters in terms of the structure of dyadic and triadic interactions, encoded in the adjacency matrix ${\bf A}$ and tensor ${\bf B}$. It does not require knowledge of the individual oscillator frequencies, but it is unable to capture finite size effects arising from the specific realization of these frequencies.
\newpage
We note that the self-averaging assumption underlying the FDA is expected to hold most accurately for nodes with large degree, for which the contribution of neighboring oscillators can be approximated by an average over the frequency distribution. However, in heterogeneous networks, high-degree nodes can strongly influence the dynamics of their neighbors, inducing correlations between local mean fields and oscillator frequencies. These correlations violate the independence assumptions required for the FDA and can lead to discrepancies, particularly near the onset of synchronization or in networks with pronounced degree heterogeneity.

 Having the local order parameters from either direct numerical simulation of Eq.~(\ref{eq:ode_sys}), from the TAT approximation (\ref{eq:h_tat_system})-(\ref{eq:r3_tat_system}), or from the FDA (\ref{eq:h_fda_system})-(\ref{eq:r3_fda}), macroscopic synchronization can be measured from the global order parameter given by
\begin{align}
R^{(2)} = \frac{\sum_n R_n^{(2)}}{\sum_n k_n}, \label{globalR}
\end{align}

 which is $0$ when the oscillators are incoherent and $1$ when they are synchronized.
 %The results we obtain here are the same as in Sabhahit (2024) [cite]. The procedure for approximation from both TAT and FDA is to calculate $R^{(2)}_n$ and $R^{(3)}_n$ to get $H_n$ and then perform the same process recursively until the order parameter value converges [cite].

\subsection{Heterogeneous Mean-Field Approximation}\label{HMF} \

If we introduce a heterogeneous mean-field approximation~\cite{Bick2023,Satorras2015,Dorogovtsev2008}, our results reduce to those obtained using the Ott-Antonsen ansatz in Ref.~\cite{Adhikari2023}. In this mean-field approximation, we assume that the local order parameters of node $n$ depend only on its weighted nodal degrees. The dyadic and triadic weighted degrees $k_n$ and $q_n$ of node $n$ are given, respectively, by
\begin{align}
k_n &= \sum_{m=1}^N \boldsymbol{A}_{nm}, \\
q_n &= \sum_{m,j}^N \boldsymbol{B}_{nmj}.
\end{align}

%In the following, we will consider the special case $q_n = k_n$ treated in Ref.~\cite{Adhikari2023} in order to show how we recover their results in the heterogeneous mean-field approximation. 
One approach to derive a  heterogeneous mean-field approximation is to assume that the network and hypergraphs have the averaged structure (see Refs.~\cite{Dorogovtsev2008, Landry2022}) 
\begin{align}
\boldsymbol{A}_{nm} &= \frac{k_n k_m}{N\langle k \rangle}\label{hetemf1},\\
\boldsymbol{B}_{nmj} &= \frac{2q_n q_m q_j}{(N\langle q \rangle)^2}\label{hetemf2},
\end{align}
where the brackets $\langle \cdot \rangle$ indicate an average over nodes, $\langle x \rangle \equiv \sum^{N}_{n=1} x_n/N$, and that the local order parameters are proportional to the weighted degrees, i.e.,
\begin{align}
    R^{(2)}_n &\approx \alpha k_n, \label{eq:alpha}\\
    R^{(3)}_n &\approx \beta q_n,\label{eq:beta}
\end{align}
where $\alpha$ and $\beta$ are constants. With this assumption, the local mean field becomes $H_n = \alpha K_2 k_n + \beta K_3 q_n$.

The Ansatz in Eqs.~(\ref{eq:alpha}) and (\ref{eq:beta}) is motivated by the role of high-degree nodes in the synchronization process. Nodes with high degree receive inputs from a large number of neighbors. The incoherent contributions from drifting neighbors tend to average out, while the coherent synchronization signal accumulates. Consequently, high-degree nodes are more strongly driven by the mean field and tend to have a higher local order parameter [$R_n^{(2)}$] than low degree nodes. This assumed linear relationship [$R_n^{(2)} \propto k_n$, $R^{(3)}_n \propto q_n$] effectively captures this structural dependence. However, it is well known that this approximation breaks down for highly heterogeneous networks \cite{Castellano2010,Goltsev2012}.

Substituting (\ref{eq:alpha}) and (\ref{eq:beta}) into (\ref{eq:h_fda_system})-(\ref{eq:r3_fda}) and summing over $n$, we find the closed system of equations for $\alpha$ and $\beta$
\begin{align}
    \alpha  &= \frac{1}{N \langle k \rangle}\sum_m k_m \frac{\sqrt{(K_2 \alpha k_m + \beta K_3 q_m)^2+1}-1}{\alpha K_2 k_m + \beta K_3 q_m}\label{eq:hmf1}\\
       \beta  &= 2\alpha \frac{1}{N \langle q \rangle}\sum_{m} q_m\left(\frac{\sqrt{(K_2 \alpha k_m + \beta K_3 q_m)^2+1}-1}{\alpha K_2 k_m + \beta K_3 q_m}\right)^2,\label{eq:hmf2}
\end{align}
from which the global order parameter can be found as
\begin{align}
R^{(2)} = \frac{\sum_n R_n^{(2)}}{\sum_n k_n} = \alpha.
\end{align}

This formulation agrees with the case $k_n = q_n$ treated in Eqs.~(35)-(37) in Ref.~\cite{Adhikari2023} with the identification $\alpha = U_1$, $\beta = 2\alpha U_2$, and $b_n = [\sqrt{(K_2 \alpha k_n + \beta K_3 k_n)^2+1}-1][\alpha K_2 k_n + \beta K_3 k_n]^{-1}$.

\subsection{Summary of Approximations} \

In this Section we derived three alternative approximations to the order parameters associated with Eq.~(\ref{eq:ode_sys}): the TAT, the FDA, and the HMF. Table~\ref{tab:approximations} summarizes the main differences between these approximations.

\begin{table}[H]
\caption{Summary of the three approximations to the order parameters associated with Eq.~(\ref{eq:ode_sys}), indicating whether each retains the realization of the intrinsic frequencies and of the hypergraph.}
\label{tab:approximations}
\centering
\renewcommand{\arraystretch}{1.5}
\setlength{\tabcolsep}{8pt}
\begin{tabular}{|c|c|c|}
\hline
{\normalsize Approximation} & {\normalsize Keeps frequency realization?} & {\normalsize Keeps hypergraph realization?}\\
\hline
{\normalsize TAT} & {\normalsize Yes} & {\normalsize Yes} \\
\hline
{\normalsize FDA} & {\normalsize No} & {\normalsize Yes }\\
\hline
{\normalsize HMF }& {\normalsize No }& {\normalsize No }\\
\hline
\end{tabular}
\end{table}
\section{Onset of Bistability} \label{sec:PT_BI} \
 
One of the most notable phenomena that the introduction of higher-order interactions brings to oscillator synchronization is that of explosive synchronization, bistability, and hysteresis~\cite{Skardal2020,Adhikari2023,Battiston2020}. In particular, it has been shown that for the Kuramoto model with higher-order interactions in Eq.~(\ref{eq:ode_sys}), the transition to synchronization becomes discontinuous for $K_3$ larger than a critical value $K_3^c$. The value of this critical triadic coupling strength has been determined for the all-to-all case ~\cite{Skardal2020} and approximated for dense hypergraphs under the heterogeneous mean-field approximation \cite{Adhikari2023}. For the latter, particularly when dyadic and triadic degrees are equal ($k_n=q_n$), the coupling strength was determined to be
\begin{align}
K_3^{\text{c}} \approx \frac{\langle k^4\rangle \langle k \rangle^2}{\langle k^2\rangle^2 \langle k^3 \rangle}.\label{oldk3}
\end{align} 

As in the dyadic-only case ~\cite{Restrepo2005}, we expect this approximation to break down when the hypergraph degree distribution is heterogeneous. In this Section, we derive an improved estimate for $K_3^c$ based on the FDA by performing a perturbative expansion of Eqs.~(\ref{eq:h_fda_system})-(\ref{eq:r3_fda}) for small $R_n^{(2)}$, $R_n^{(3)}$. For this, we let
\begin{align}
R_n^{(2)} = \epsilon \hat R_n^{(2)} + \epsilon^3 \tilde R_n^{(2)} + \dots,\\
R_n^{(3)} = \epsilon \hat R_n^{(3)} + \epsilon^3 \tilde R_n^{(3)} + \dots,
\end{align}
where $\epsilon$ is an expansion parameter that we formally consider small. First, we consider the lowest order in $\epsilon$. Inserting these expressions in Eqs.~(\ref{eq:h_fda_system})-(\ref{eq:r3_fda}), expanding in series, and neglecting terms of order $\epsilon^3$, we find
\begin{align}
\hat R_n^{(2)} &=\frac{K_2}{2} \sum^N_{m=1} \boldsymbol{A}_{nm} \hat R_m^{(2)}, \label{eq:R2hat}\\
\hat R_n^{(3)} &= 0.
\end{align}

Therefore, the dyadic coupling strength for the onset of synchrony is 

%In this section, we employ the solution from FDA to find the critical point for the phase transition. We estimate the order parameter at the transition point where $H_n$ is non-zero but infinitesimal, $H_n \ll 1$. Therefore, the results from FDA reduce to the following:
%\begin{align}
% 	\frac{\sqrt{H_n^2+1}-1}{H_n} &\approx \frac{H_n}{2} - \frac{H_n^3}{8} + %\mathcal{O}(H_n^5), \label{eq:appx_R2}
%\end{align}
%and then Eq. \ref{eq:appx_R2} helps reduce $H_n$ to a simple form as:
%\begin{align}
% 	H_n &= K^c_2 R^{(2)}_n + K_3 R^{(3)}_n \label{eq:Hn_cc}\\ 
%    H_n &\approx K^c_2 \sum_m \boldsymbol{A}_{nm} \frac{\sqrt{H_m^2+1}-1}{H_m} + K_3 \sum_{m,j}\boldsymbol{B}_{nmj} \left(\frac{\sqrt{H_m^2+1}-1}{H_m}\right)^2 \left( \frac{\sqrt{H_j^2+1}-1}{H_j} \right) \\
%    H_n &\approx K^c_2 \sum_m \boldsymbol{A}_{nm} \frac{H_m}{2} \label{eq:k2_c}
%\end{align}
%From Eq. \ref{eq:k2_c}, we observe that this linear equation can be effectively solved using eigenvalue methods. Thus, we find that:
\begin{align}
  	K^c_2 = \frac{2}{\lambda} \label{eq:eig_val_A},
\end{align}
where $\lambda$ is the Perron--Frobenius (principal) eigenvalue of the dyadic adjacency matrix $\boldsymbol{A}$. Rewriting Eq.~\eqref{eq:R2hat} in vector form yields the eigenvalue problem $\boldsymbol{A} \hat{\boldsymbol{R}}^{(2)} = (2 / K_2) \hat{\boldsymbol{R}}^{(2)}$. Since $\boldsymbol{A}$ has non-negative entries and the local order parameters are non-negative, the Perron--Frobenius theorem implies (for a connected network) that the only eigenvector with non-negative entries is the one associated with the largest eigenvalue $\lambda$. Therefore, a non-trivial, non-negative solution first emerges when $2 / K_2$ equals the largest eigenvalue $\lambda$, yielding $K_2^c = 2 / \lambda$. The onset of synchrony agrees with that obtained for the case of dyadic interactions only~\cite{Restrepo2005}. Furthermore, from Eq.~(\ref{eq:R2hat}) we see that to leading order the local dyadic order parameter is proportional to the right eigenvector $\boldsymbol{U}$ of matrix $\boldsymbol{A}$.
\begin{align}
  	\hat R_n^{(2)} = C U_n,
\end{align}
with the constant $C$ being undetermined at the lowest order of the perturbation.
  
To investigate the system's behavior near the critical point $K_2=K_2^c$, we introduce a perturbation to the coupling strength, i.e., we let $K_2 = K^c_2 + \epsilon^2 \delta K$. Inserting 
\begin{align}
K_2 = K^c_2 + \epsilon^2 \delta K, \hspace{1cm} R_n^{(2)} = \epsilon CU_n + \epsilon^3 \tilde R_n^{(2)}, \hspace{1cm} R_n^{(3)} = \epsilon^3 \tilde R_n^{(3)},
\end{align}
into Eqs.~(\ref{eq:h_fda_system})-(\ref{eq:r3_fda}) and expanding them to cubic order in $\epsilon$, we find
\begin{align}
%H_n &= \epsilon \frac{2}{\lambda} C U_n + \epsilon^3\left( \frac{2}{\lambda} \tilde R_n^{(2)} + C U_n \delta K + K_3 \tilde R_n^{(3)}\right),\\
\epsilon C U_n + \epsilon^3 \tilde R_n^{(2)} &= \sum_m \boldsymbol{A}_{nm} \left( \frac{1}{\lambda} \epsilon C U_m + \frac{\epsilon^3}{\lambda} \tilde R_m^{(2)} +\frac{\epsilon^3}{2} C U_m \delta K +\frac{\epsilon^3}{2} K_3 \tilde R_m^{(3)}  - \frac{1}{\lambda^3}\epsilon^3 C^3 U_m^3\right),\\
\tilde R_n^{(3)} &= \sum_{m,j} B_{nmj}\frac{C^3}{\lambda^3}U_m^2 U_j.
\end{align}
Using the fact that $\sum_m \boldsymbol{A}_{nm}U_m = \lambda U_n$ and simplifying, we get
\begin{align}
 \tilde R_n^{(2)} &= \frac{1}{\lambda} \sum_m \boldsymbol{A}_{nm}  \tilde R_m^{(2)} +\frac{1}{2} \lambda C U_n \delta K + \frac{K_3}{2}   \sum_m \boldsymbol{A}_{nm} \tilde{R}_m^{(3)}  - \frac{C^3}{\lambda^3}  \sum_m \boldsymbol{A}_{nm} U_m^3,\label{messy}\\
\tilde R_n^{(3)} &= \frac{C^3}{\lambda^3}\sum_{m,j} B_{nmj}U_m^2 U_j.
\end{align}\

In order to solve for the amplitude $C$, we eliminate the unknown perturbation $\tilde R_n^{(2)}$ by multiplying by the $n^{\text{th}}$ entry of the left eigenvector of the adjacency matrix, $V_n$, and summing over $n$, where the left eigenvector satisfies $\lambda V_m = \sum_n V_n \boldsymbol{A}_{nm}$.
Using $\sum_n V_n \boldsymbol{A}_{nm} = \lambda V_m$, the first term on the right-hand side of Eq.~(\ref{messy}) becomes $\frac{1}{\lambda}\sum_{n,m} V_n\boldsymbol{A}_{nm}\tilde R_m^{(2)} = \sum_m V_m \tilde R^{(2)}_m$, which cancels exactly with the left-hand side, $\sum_n V_n \tilde R^{(2)}_n$. Note that this does not require $\sum_n V_n \tilde R^{(2)}_n$ to vanish: it is the standard solvability condition, which follows from the fact that the operator $\boldsymbol{I} - \boldsymbol{A}/\lambda$ acting on $\tilde{\boldsymbol{R}}^{(2)}$ is singular, with left null vector $\boldsymbol{V}$. The remaining terms give
\begin{align}
\frac{\lambda \delta K C}{2}\sum_n U_n V_n + \frac{K_3 \lambda}{2}\sum_m V_m \tilde R^{(3)}_m - \frac{C^3}{\lambda^2}\sum_m V_m U_m^3 = 0.
\end{align}
Inserting the expression for $\tilde R^{(3)}_m$ and solving for $C$ leads to the following expression for the square of the amplitude $C$:
\begin{align}
    C^2 = \frac{\lambda^3 \delta K \sum_n U_n V_n}{2 \sum_{m} V_m U_m^3 - K_3 \sum_{n,m,j} \boldsymbol{B}_{nmj}  V_n U_m^2 U_j}. \label{eq:Csq}
\end{align}

This equation allows us to identify the critical triadic coupling strength $K_3^c$ for the onset of bistability. In a bistable regime, a coherent state (characterized by $C>0$) may persist even when the perturbation $\delta K$ is negative. For the amplitude $C$ to remain real and non-zero, the denominator in Eq.~(\ref{eq:Csq}) must share the sign of the numerator. Therefore, the critical value of $K_3$ is given by
\begin{align}
   K_3^c = \frac{2 \sum_{m} V_m U_m^3}{\sum_{n,m,j} \boldsymbol{B}_{nmj}  V_n U_m^2 U_j},\label{eq:K3c_eq}
   %= K_2^c\frac{\sum_{n,m,j}V_n \boldsymbol{A}_{nm} \boldsymbol{A}_{mj} U_m^2U_j}{\lambda \sum_{n,m,j}V_n\boldsymbol{B}_{nmj}U_m^2U_j} 
\end{align}
with bistability occurring for $K_3 > K_3^c$. Equation~(\ref{eq:K3c_eq}), one of our main results, explicitly quantifies the dependence of the critical point $K_3^c$ on the network's spectral properties. In particular, it shows that the onset of bistability depends on how the eigenvectors of the adjacency matrix encoding dyadic connections are correlated with the structure of the triadic interactions. For a given dyadic interaction network, the value of $K_3^c$ is minimized (bistability is promoted), if the triadic interactions are chosen to maximize the sum
\begin{align}
\sum_{n,m,j}\boldsymbol{B}_{nmj}V_n U_m^2 U_j.
\end{align}

\begin{figure}[t]

    \centering
    
    \begin{subfigure}{0.4\textwidth}
        \centering
        \includegraphics[width=\linewidth]{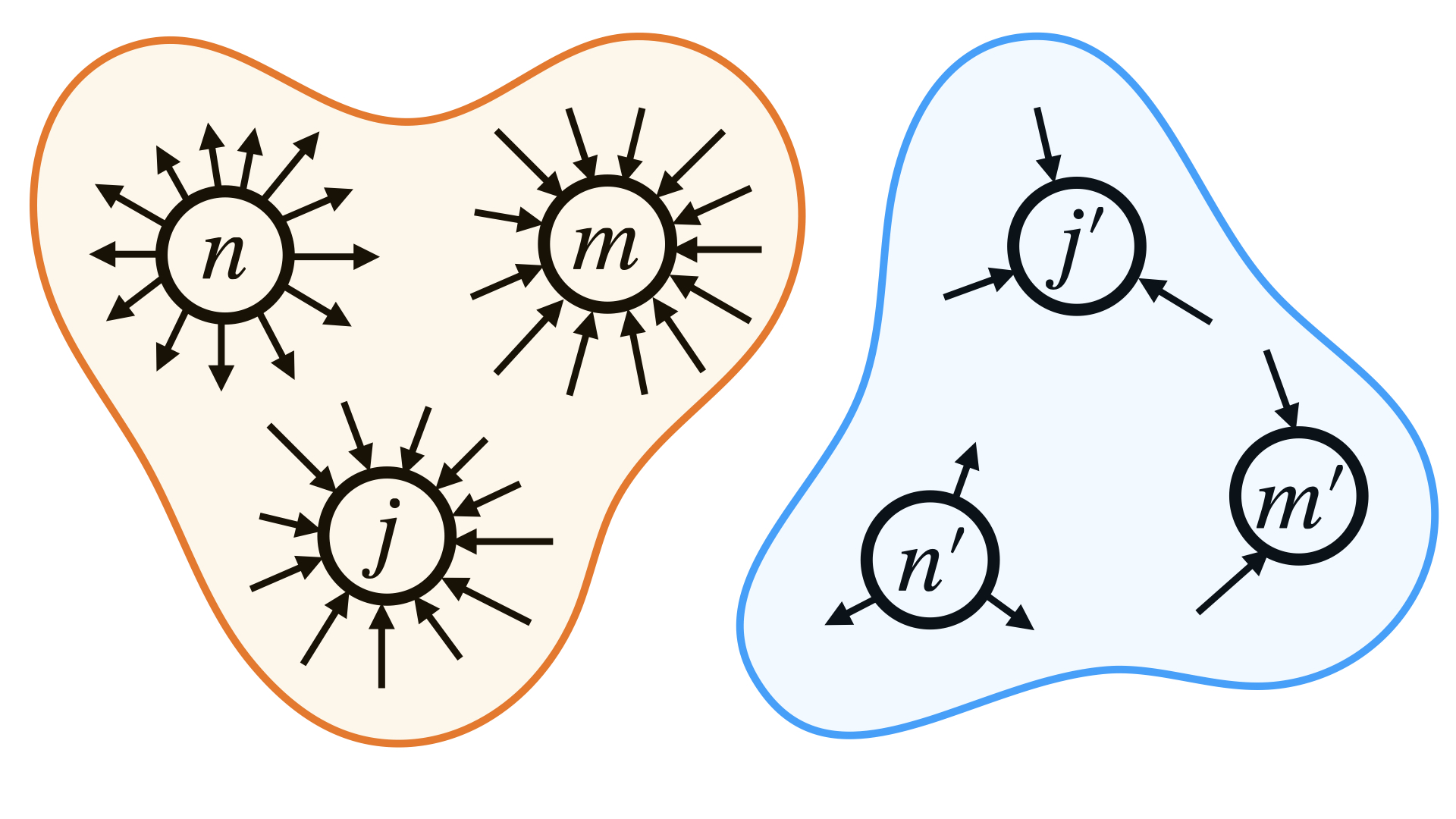}
        \caption{Lower $K^c_3$.}
        \label{fig:lower_k3}
    \end{subfigure}
    \hfill
    \begin{subfigure}{0.4\textwidth}
        \centering
        \includegraphics[width=\linewidth]{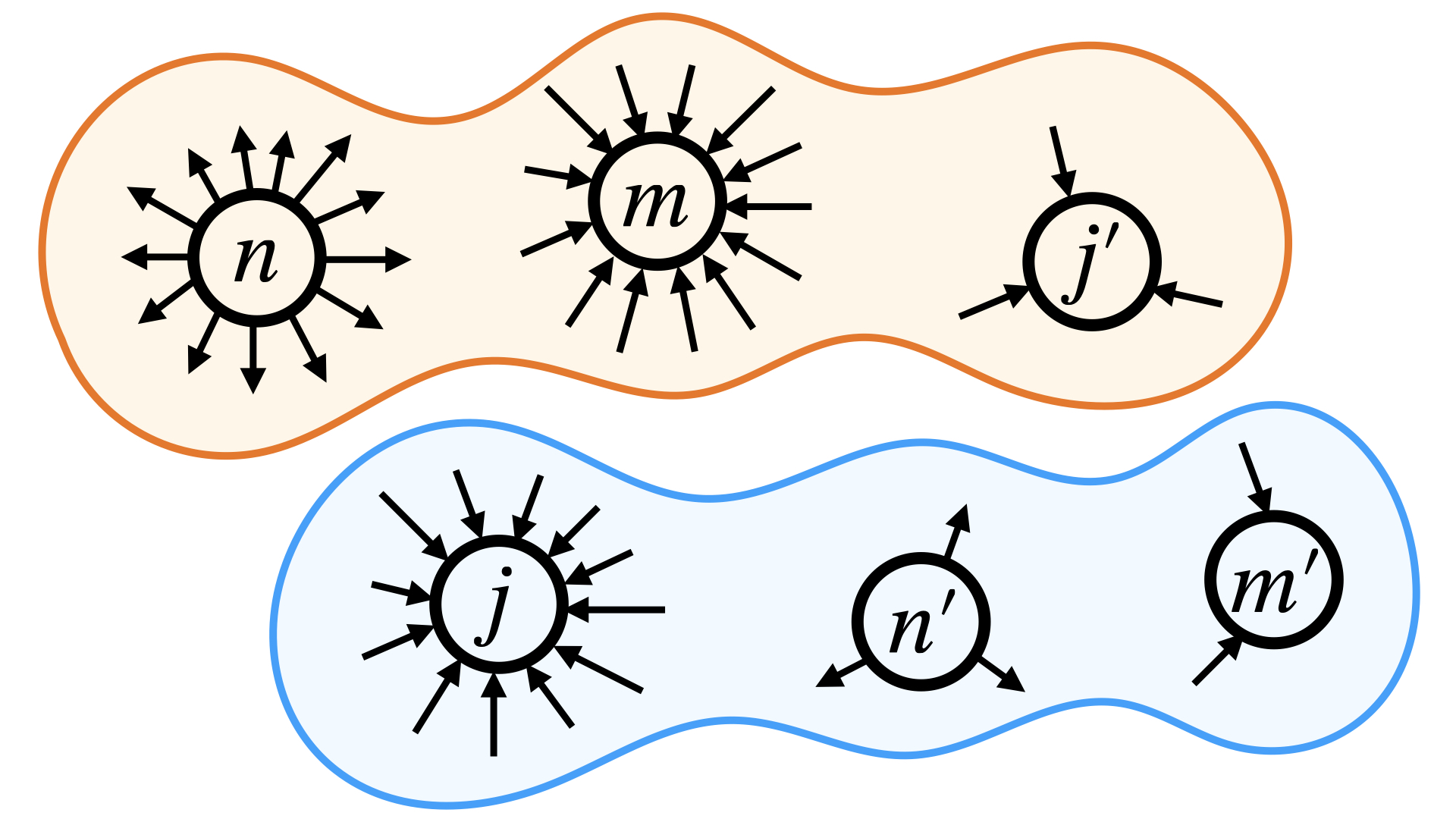}
        \caption{Higher $K^c_3$.}
        \label{fig:higher_k3}
    \end{subfigure}
\par\bigskip
    \begin{subfigure}{0.4\textwidth}
        \centering
        \includegraphics[width=\linewidth]{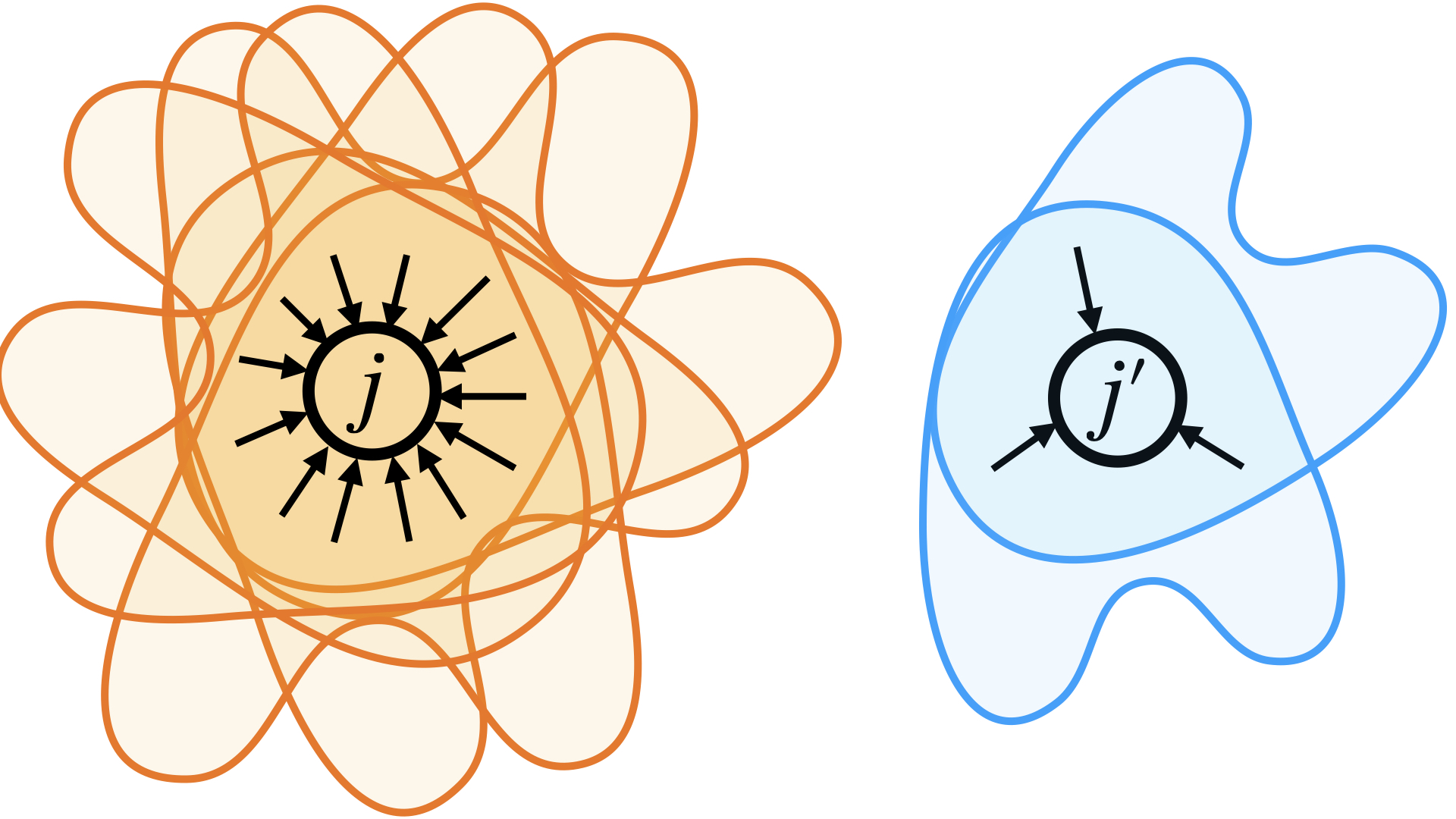}
        \caption{Lower $K^c_3$.}
        \label{fig:corr_kq}
    \end{subfigure}
    \hfill
    \begin{subfigure}{0.4\textwidth}
        \centering
        \includegraphics[width=\linewidth]{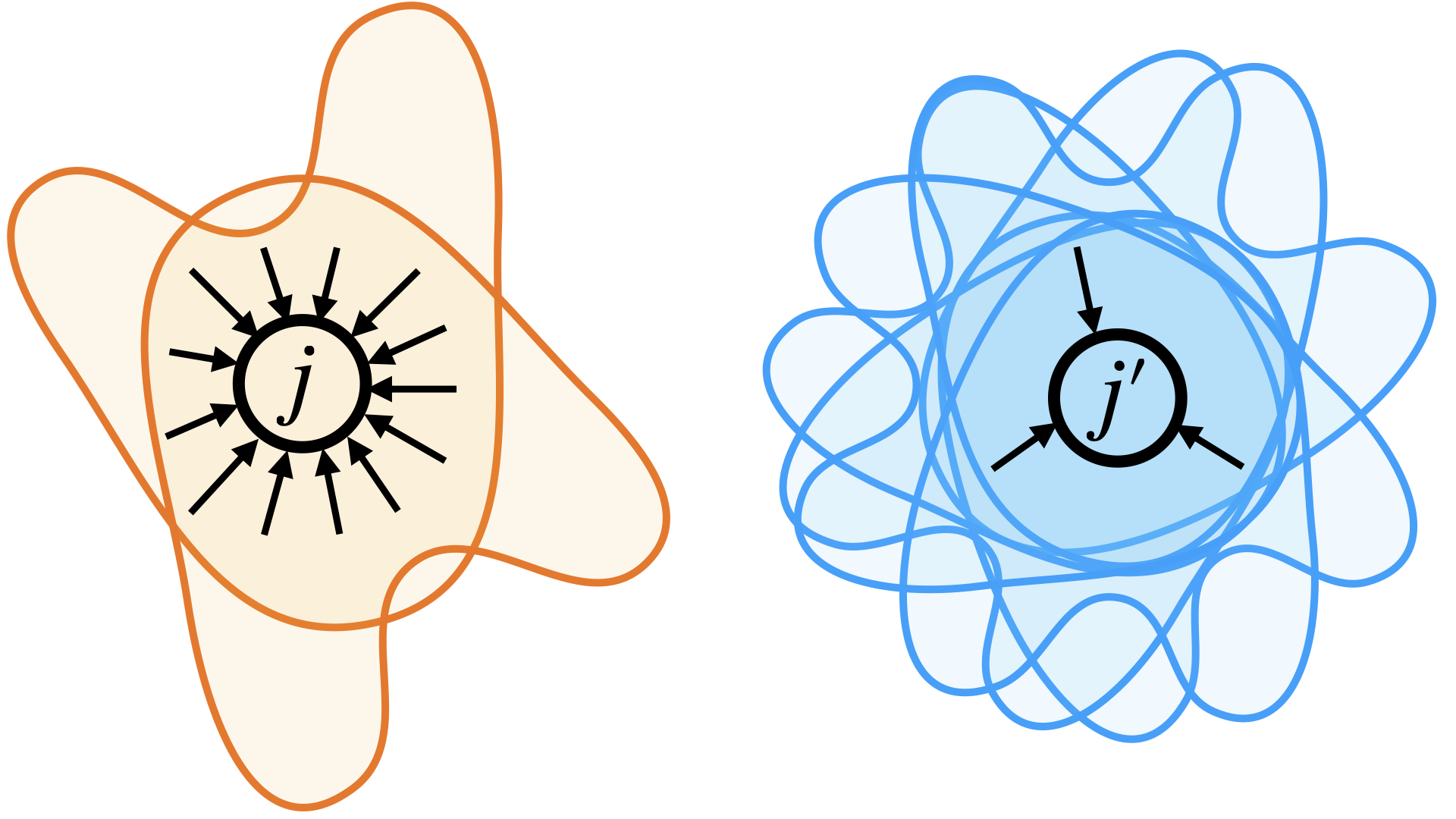}
        \caption{Higher $K^c_3$.}
        \label{fig:anti_corr_kq}
    \end{subfigure}

    \caption{Effect of hypergraph structure on $K_3^c$. (Top panels) Nodes $\{n,m,j\}$ have high values of $V_n$, $U_m$, $U_j$, respectively (e.g., by having large dyadic degrees), while nodes $\{n',m',j'\}$ have low values of $V_{n'}$, $U_{m'}$, $U_{j'}$. Configurations where hyperedges couple oscillator triplets $\{n,m,j\}$ and $\{n',m',j'\}$ (a) have lower values of $K_3^c$ than those where those hyperedges couple other triplets, $\{n,m,j'\}$ and $\{n',m',j\}$ (b). (Bottom panels) Visualizations of hypergraphs where triadic and dyadic degrees are either correlated (c) or anti-correlated (d).}
    \label{fig:corr_swap}
\end{figure}

For an undirected network, where $U_n = V_n$, bistability is promoted when triadic interactions preferentially couple nodes with large eigenvector entries $U_n$. Since these entries are often correlated with dyadic degree, hyperedges connecting highly connected nodes tend to lower $K_3^c$. This is illustrated schematically in Fig.~1: the hypergraph in panel (a), where hyperedges connect nodes with similarly large or similarly small eigenvector entries, has a lower value of $K_3^c$ than the hypergraph in panel (b), where nodes with high and low eigenvector entries are mixed.

The intuition behind this result is as follows. Near the onset of synchronization, the dyadic coupling saturates the growth of the order parameter [the term proportional to $\sum_m V_m U_m^3$ in Eq.~(\ref{eq:Csq})], which by itself results in a continuous transition. The triadic coupling, on the other hand, is cubic in the order parameter and provides a positive nonlinear feedback [the term proportional to $K_3 \sum_{n,m,j}\boldsymbol{B}_{nmj} V_n U_m^2 U_j$]. Bistability appears when this feedback overcomes the dyadic saturation. Since the entries $U_n$ and $V_n$ measure how strongly each node participates in and responds to the growing synchronization mode, the triadic feedback is strongest when hyperedges join nodes that are themselves strongly coherent and influential, as in Fig.~\ref{fig:lower_k3}, and weakest when they join such nodes to weakly coherent ones, as in Fig.~\ref{fig:higher_k3}.

In the heterogeneous mean-field approximation [cf. Eqs.~(\ref{hetemf1})-(\ref{hetemf2})] $K_3^c$ reduces to 
\begin{align}
K_3^{\text{HMF}} = \frac{\langle k^4\rangle \langle q\rangle^2}{\langle q k\rangle^2 \langle q k^2\rangle}. \label{newk3}
\end{align} 

This expression further reduces to (\ref{oldk3}) if $k_n = q_n$. Importantly, Eq.~(\ref{newk3}) shows how the coupling across interactions of different orders affects the dynamics. In particular, when the number of dyadic and triadic interactions per node are uncorrelated, we have $\langle q k \rangle = \langle q \rangle \langle k \rangle$ and $\langle q k^2 \rangle = \langle q \rangle \langle k^2 \rangle$, and the critical coupling strength reduces to 
\begin{align}
K_3^{\text{HMF}}  \to \frac{\langle k^4\rangle}{\langle k\rangle^2 \langle  k^2\rangle\langle  q\rangle}, \label{newk3unco}
\end{align}
which is independent of heterogeneity in the triadic degree distribution. On the other hand, when the dyadic and triadic degrees are correlated (the extreme case being $q_n = k_n$) then the terms in the denominator of (\ref{newk3}) are larger and, consequently, $K_3^{\text{HMF}}$ is smaller. This is illustrated  in Figs.~\ref{fig:corr_kq}-~\ref{fig:anti_corr_kq}. The left panel, Fig.~\ref{fig:corr_kq}, shows a situation where dyadic and triadic degrees are correlated, while they are anti-correlated in the right panel, Fig.~\ref{fig:anti_corr_kq}. Finally, we note that heterogeneity in the dyadic degree distribution increases $K_3^c$, due to the term $\langle k^4 \rangle$ in the numerator of (\ref{newk3}).
Table~\ref{tab:structure_effects} summarizes how the structure of dyadic and triadic interactions affects the onset of bistability.
\begin{table}[H]
\caption{Effect of the structure of dyadic and triadic interactions on the critical triadic coupling strength $K_3^c$ for the onset of bistability, and the approximation from which each effect is derived.}
\label{tab:structure_effects}
\centering
\renewcommand{\arraystretch}{1.5}
\setlength{\tabcolsep}{8pt}
\begin{tabular}{ |c|c|c| } 
 \hline
Feature & Effect & Approximation used \\ 
 \hline
Triplets with high eigenvector entries coupled & lower $K_3^c$ & FDA \\ 
 \hline
Triadic and dyadic degrees correlated & lower $K_3^c$ & FDA and HMF \\ 
 \hline
 Dyadic interactions heterogeneous & higher $K_3^c$ & FDA and HMF \\ 
 \hline
\end{tabular}
\end{table}

\section{Summary of Assumptions}\label{assumptions} \

In the previous sections we derived various theoretical characterizations of the macroscopic order parameter as a function of the dyadic and triadic coupling strengths. However, these theoretical approximations came at the cost of making various assumptions, which we summarize and discuss in this Section. 

\begin{enumerate}
\item The first assumption, introduced after Eq.~(\ref{eq:hn}), is that there is a solution such that $H_n$ and $\psi_n$ are constants. This assumption is analogous to the self-consistent assumption Kuramoto made in solving the all-to-all model \cite{KMreview, Kuramoto,Strogatz2000}. In our case, however, this assumption is non-trivial, because the complex number $H_n e^{i\psi_n}$ is defined in terms of finite sums of oscillating terms. When there are few terms in these sums, finite-size fluctuations may not be negligible. In Ref.~\cite{Restrepo2005}, the effect of these fluctuations was studied for the dyadic Kuramoto model and it was found that they can delay the transition to synchronization. We do not conduct such an analysis here, but instead assume that the number of terms in the sums (\ref{eq:local_R2_def})-(\ref{eq:local_R3_def}) is large.
\item Another important assumption is that the constant phase angle $\psi_n$ is the same for all oscillators. This effectively assumes a single synchronization cluster, instead of multiple clusters synchronizing with different phases, such as would occur in hypergraphs with community or spatial structure. While it would be possible to extend our framework to treat various communities, we leave this extension for future work.
\item The next assumption is that the frequencies $\omega_n$ and local order parameters $H_n$ are uncorrelated, introduced after Eq.~(\ref{eq:each_term_R2}). As discussed previously, we expect this approximation to be most accurate for dense networks that are not too heterogeneous.
\item We also assumed, as is typically done, that the frequency distribution $g(\omega)$ is symmetric. Using assumption 3 above and the assumption that the number of terms in the sum is large, the imaginary terms and the contribution from drifting oscillators in Eqs.~(\ref{eq:each_term_R2}) and (\ref{eq:R3_LL})-(\ref{eq:R3_DD})  can be neglected.
\item The last assumption required to derive the TAT is that  pair correlations can be neglected. As discussed in Ref.~\cite{Adhikari2023}, we expect this approximation to be good for dense networks or when close to full synchronization or incoherence. 

\item In order to derive the FDA from the TAT, we approximate the sums in Eqs.~(\ref{eq:r2_tat_system})-(\ref{eq:r3_tat_system}) by their averages over the frequency distribution $g(\omega)$. This requires a self-averaging assumption, whereby each node interacts with a sufficiently large number of neighbors so that fluctuations due to individual frequency realizations can be neglected.

\item The derivation of the critical triadic coupling strength $K_3^c$ in Eq.~(\ref{eq:K3c_eq}) is based on a perturbative expansion about the critical point $K_2=K_2^c$ where the order parameters align with the Perron-Frobenius eigenvectors ${\bf U}$, ${\bf V}$. The derived value of $K_3^c$, therefore, relies on synchronization being driven by a dominant mode, and would break down in situations where multiple modes compete.

\item All the results above are derived for the specific triadic coupling in Eq.~(\ref{eq:ode_sys}) without phase lags. With a phase lag, drifting oscillators contribute to $R^{(2)}_n$ and the self-consistent equations must be modified, and for other forms of triadic coupling the expressions for $R^{(3)}_n$ and for $K_3^c$ change.

\item Finally, the HMF approximation further assumes that all nodes with the same degree behave statistically the same, and that the local order parameters are proportional to the degree, as postulated and discussed in and after Eqs.~(\ref{eq:alpha})-(\ref{eq:beta}). This assumption excludes networks with correlations (such as degree assortativity), but the theory could be extended by considering probabilities of connection that depend on nodal variables.
\end{enumerate}

In summary, while our framework relies on various assumptions, most of them are satisfied for large, dense networks. The other assumptions, such as the assumption of a single synchronization cluster or the lack of degree assortativity, could potentially be relaxed. A rigorous characterization of the errors introduced by these approximations is outside the scope of this paper.

\section{Validation and Examples} \label{sec:validation} \

In this Section we validate our theoretical results with direct numerical simulations of Eq.~(\ref{eq:ode_sys}).

\subsection{Hypergraph generative model}\label{generativemodel} \

In order to generate the hypergraphs to simulate Eq.~(\ref{eq:ode_sys}), we use the Chung-Lu model \cite{Chung2002} adapted to hypergraphs, as in Refs.~\cite{Adhikari2023, Zhou2007, Stasi2014}. Given a value of $N$ and desired dyadic and triadic degree distributions, we generate target degree sequences sampled independently from these distributions, $[\hat k_1, \hat k_2,\dots, \hat k_N]$ and $[\hat q_1, \hat q_2,\dots, \hat q_N]$. We then create the hypergraph by setting $\boldsymbol{A}_{nm} = 1$ with probability $\hat k_n \hat k_m/(N \langle \hat k \rangle)$ and $\boldsymbol{A}_{nm} =0$ otherwise, and similarly $\boldsymbol{B}_{nmj} = 1$ with probability $2 \hat q_n \hat q_m \hat q_j/(N \langle \hat q \rangle)^2$ and $\boldsymbol{B}_{nmj} =0$ otherwise. The frequencies $\omega_n$ are sampled from a Lorentzian distribution with width one centered at zero, $g(\omega) = 1/[\pi(1+\omega^2)]$ (if necessary, one can redefine time to set the width at one and move to a rotating frame of coordinates to set the mean frequency at zero).
 
% We examine two distinct network architectures: (i) a relatively homogeneous graph with a degree distribution uniform in $[0.015N, 0.035N]$ (note that the mean degree is always $N/40$) and (ii) a heterogeneous graph characterized by a truncated power-law degree distribution given by
% \begin{align}
% P(k) = 
% \begin{cases} 
% C k^{-\gamma}, & k_{\text{min}} < k < k_{\text{max}}, \\
% 0, & \text{otherwise}.
% \end{cases}
% \end{align}\
% The value of $k_\text{min}$ is chosen so that, again, the mean degree is $N/40$. Values for $N$, $\gamma$ and $k_\text{max}$ will be noted separately for each simulation. Unless noted otherwise, we will choose $\hat{q}_n=\hat{k}_n$
We examine two distinct network architectures: (i) a relatively homogeneous graph with a uniform degree distribution over $[k_{\text{min}}, k_{\text{max}}]$ and (ii) a heterogeneous graph characterized by a truncated power-law degree distribution given by
\begin{align}
P(k) = \begin{cases} 
C k^{-\gamma}, & k_{\text{min}} < k < k_{\text{max}}, \\ 
0, & \text{otherwise}. 
\end{cases}
\end{align}
The specific values of $N$, the degree bounds $k_{\text{min}}$ and $k_{\text{max}}$, the mean degree $\langle k \rangle$, and the power-law exponent $\gamma$ are specified for each experiment in the corresponding figure captions. Unless noted otherwise, we choose $\hat{q}_n = \hat{k}_n$.

\subsection{Numerical methods}\label{numerical} \

In the following, we will present various Figures showing how the global order parameter $R^{(2)}$ depends on $K_2$ when $K_2$ is adiabatically increased or decreased. Direct numerical simulations of Eq.~(\ref{eq:ode_sys}) serve as the reference against which the three analytical approximations (TAT, FDA, and HMF) are compared. For each value of $K_2$, Eqs.~(\ref{eq:ode_sys}) are solved numerically using the Runge-Kutta fourth order method with timestep $\Delta t = 0.01$ for 20 time units for the sweeps in Figs.~\ref{fig:uniform_res},~\ref{fig:small_nets}, and~\ref{fig:compare_rewiring}, and for 30 time units for the bistability analysis for Figs.~\ref{fig:bistab_capped}--\ref{fig:compare_corr}, and the order parameters are computed from an average of the last 10 and 15 time units, respectively. We verified that increasing the integration time does not change the results appreciably. The dyadic coupling strength is varied in steps of $\Delta K_2 = 0.005 K_2^c$. All results shown are for a single realization of the hypergraph and of the natural frequencies; the parameters of each realization are given in the corresponding figure caption. The first approximation is the TAT. For each value of $K_2$, the TAT equations are solved numerically by using a fixed-point iteration scheme where an initial guess is inserted on the right hand side of Eqs.~(\ref{eq:r2_tat_system})-(\ref{eq:r3_tat_system}), and the resulting values are repeatedly inserted in the same equations for 200 iterations, at which point we find the values converge to a fixed point. When increasing $K_2$, the initial guesses are the values of $H_n, R_n^{(2)}$, and $R_n^{(3)}$ found for the previous value of $K_2$, and similarly when decreasing $K_2$. The second approximation is the FDA, for which a similar fixed-point iteration scheme is used using Eqs.~(\ref{eq:h_fda_system})-(\ref{eq:r3_fda}). Finally, a similar approach is also used for the HMF approximation with Eqs.~(\ref{eq:hmf1})-~(\ref{eq:hmf2}). 

\subsection{Uniform degree distribution} \

\begin{figure}[b!]
    \centering

    \begin{subfigure}{0.48\textwidth}
        \centering
        \includegraphics[width=\linewidth]{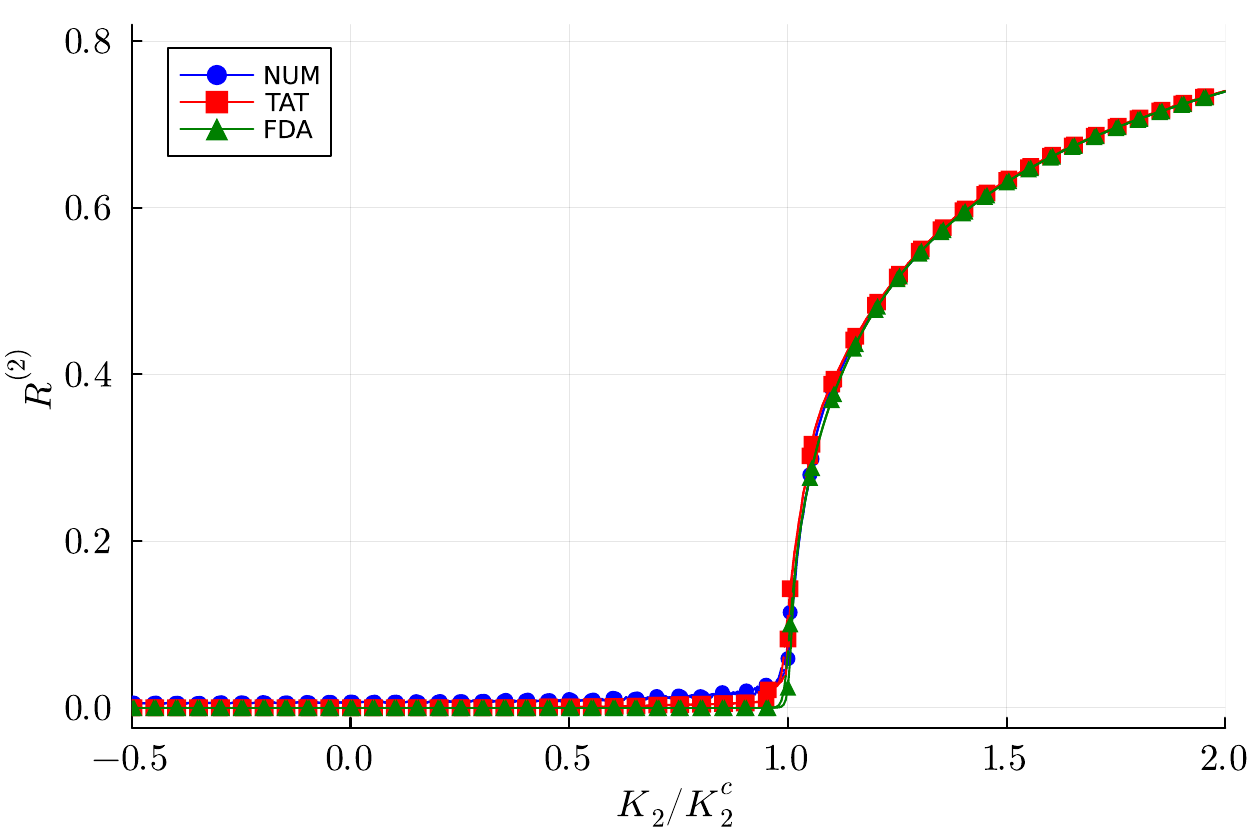}
        \caption{$K_3 < K_3^c$}
        \label{fig:res_uniform_k3lk3c}
    \end{subfigure}
    \hfill
    \begin{subfigure}{0.48\textwidth}
        \centering
        \includegraphics[width=\linewidth]{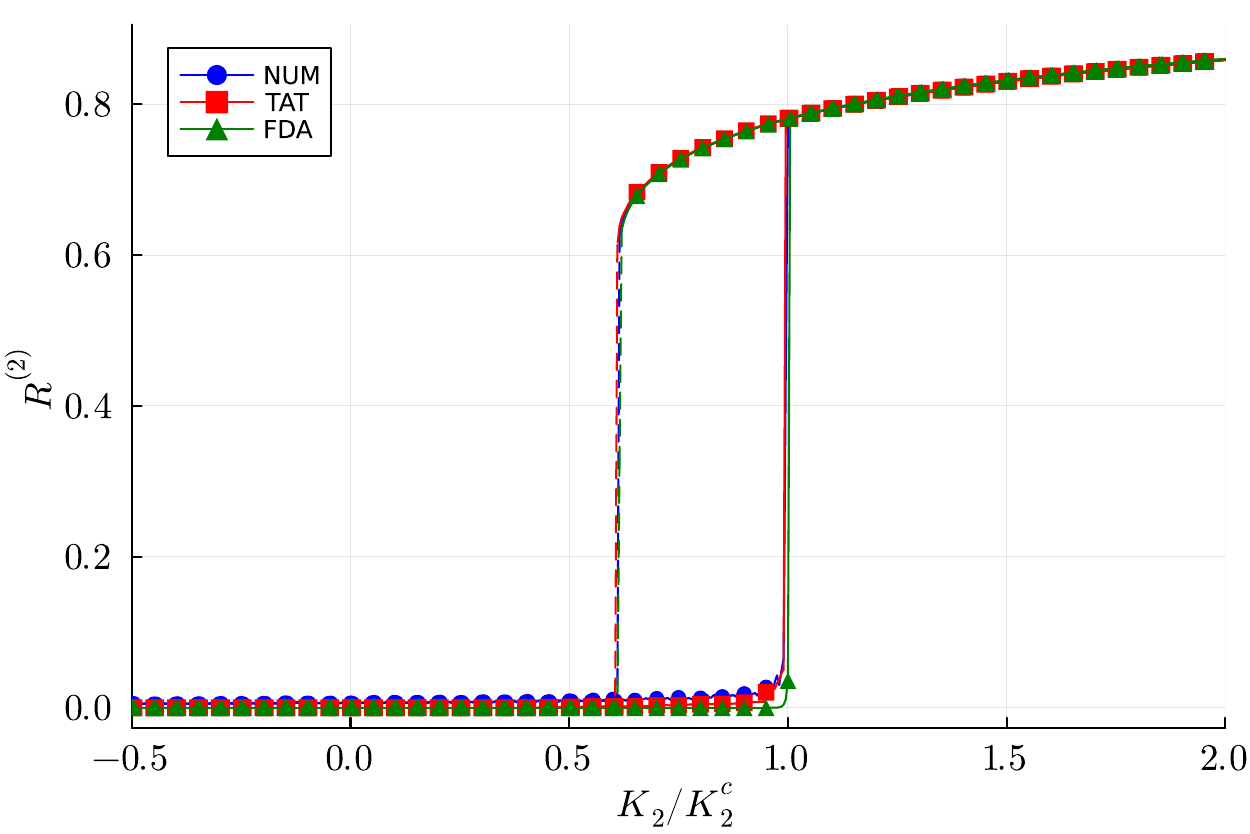}
        \caption{$K_3 > K_3^c$}
        \label{fig:res_uniform_k3gk3c}
    \end{subfigure}
    \\

   \begin{subfigure}{0.48\textwidth}
        \centering
        \includegraphics[width=\linewidth]{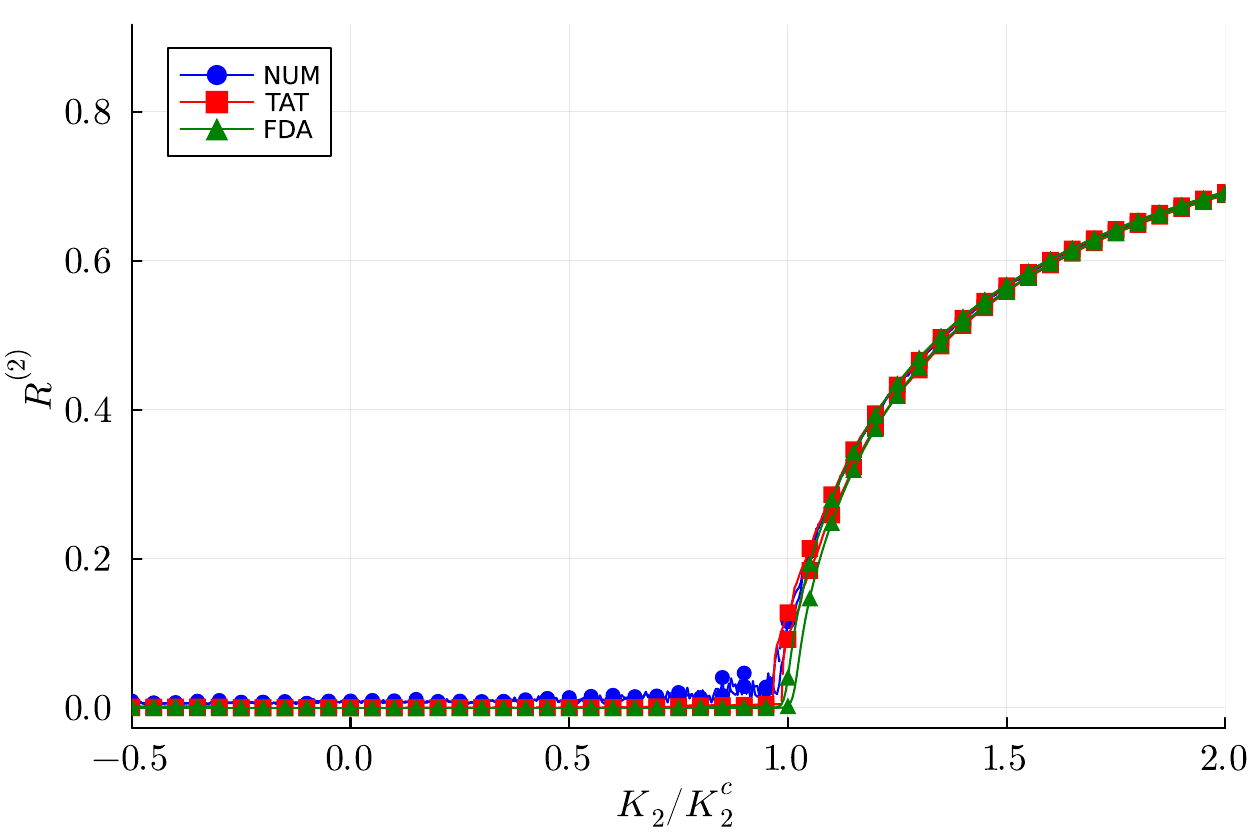}
        \caption{$K_3 < K_3^c$}
        \label{fig:res_pw_k3lk3c}
    \end{subfigure}
    \hfill
    \begin{subfigure}{0.48\textwidth}
        \centering
        \includegraphics[width=\linewidth]{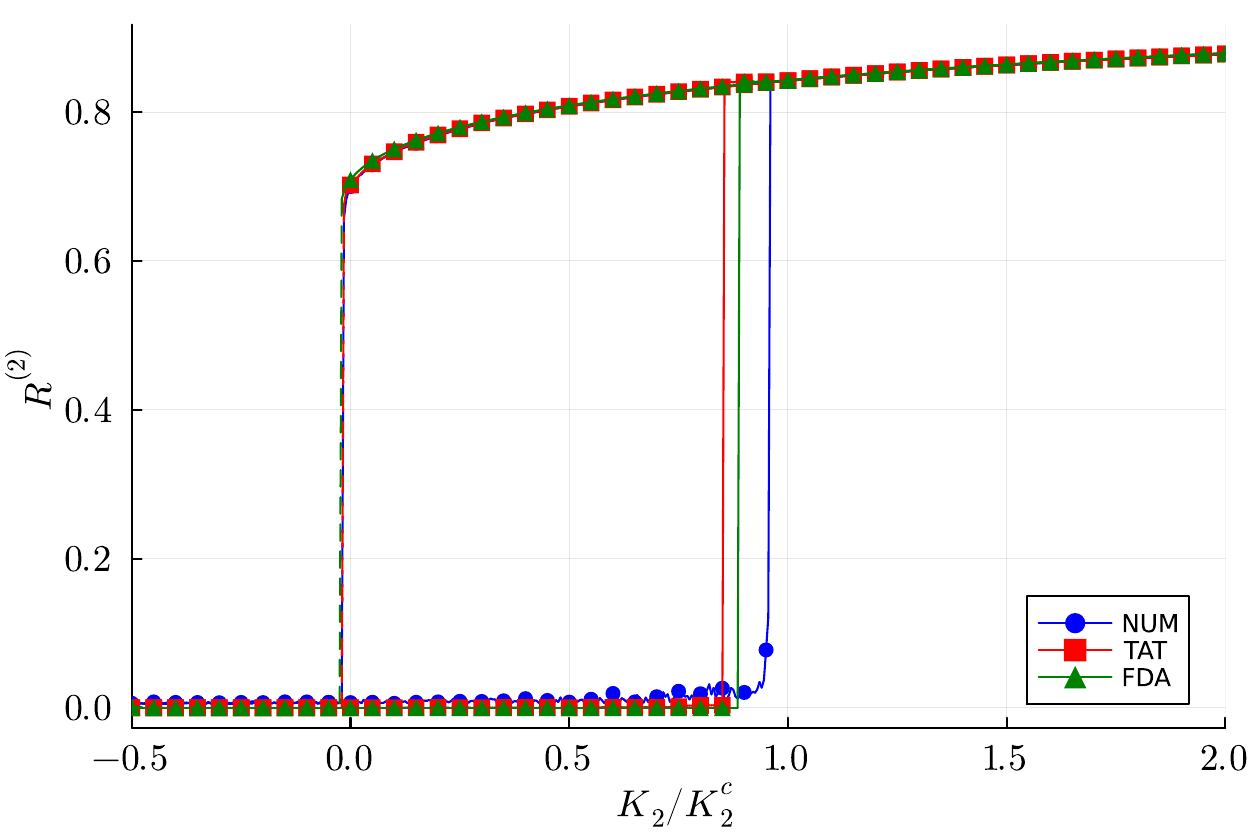}
        \caption{$K_3 > K_3^c$}
        \label{fig:res_pw_k3gk3c}
    \end{subfigure}

    \caption{Global order parameter $R^{(2)}$ as a function of $K_2/K_2^c$ for uniform (top) and power-law (bottom) degree distributions ($N=20000$, $\langle k_n \rangle = 500$; top: uniform on $[300,700]$; bottom: power law with $\gamma=3.0, k_{max}=N-1$). Panels (a) and (c) show the weak triadic coupling regime ($K_3 < K^c_3$), while (b) and (d) show the strong triadic coupling regime ($K_3 > K^c_3$). Solid and dashed lines denote forward and backward parameter sweeps, respectively.}
    \label{fig:uniform_res}
\end{figure}

We begin by analyzing the synchronization transition on a network with $N = 20000$ nodes and a uniform degree distribution for both dyadic and triadic degrees, where we additionally assume that $\hat k_n = \hat q_n$ (see the {\it correlated case} in Ref.~\cite{Adhikari2023}). The top panels of Figure \ref{fig:uniform_res} show the global order parameter $R^{(2)}$ as a function of the normalized dyadic coupling strength $K_2/K_2^c$ for $K_3 < K_3^c$ (left) and $K_3 > K_3^c$ (right). The blue curves (circles) show the order parameter obtained from direct numerical simulation of Eq.~(\ref{eq:ode_sys}), the red curves (squares) show the TAT prediction, and the green curves (triangles) show the FDA prediction. Solid and dashed lines correspond to  adiabatically increasing and decreasing $K_2$, respectively. 

In the weak triadic coupling regime ($K_3 = 0.001039 < K_3^c = 0.002078$, Fig. \ref{fig:res_uniform_k3lk3c}), the system exhibits a continuous second-order phase transition. Both the Time-Averaged Theory (TAT) and the Frequency Distribution Approximation (FDA) show excellent agreement with the numerical simulations, accurately predicting the onset of synchronization at $K_2 = K_2^c$. (The HMF, not shown, also agrees well with the simulations.)

In the strong triadic coupling regime ($K_3 = 0.005195 > K_3^c=0.002078$, Fig. \ref{fig:res_uniform_k3gk3c}), the transition becomes abrupt and first-order, with a significant hysteresis loop. The theoretical predictions successfully reproduce the bistable region where incoherent and coherent states coexist. Notably, while both approximations capture the global behavior, the TAT method provides a more accurate description of the finite-size fluctuations observed in the numerical data (note, in particular, the small increase in the order parameter for $K_2$ slightly smaller than $K_2^c$). In contrast, the FDA yields a smooth mean-field curve that represents the thermodynamic limit. This distinction suggests that the TAT is better suited for finite systems, even in structurally homogeneous networks. This will be demonstrated further below for smaller networks.

\subsection{Power-law degree distribution} \

Having validated our framework on homogeneous networks, we next examine its performance on heterogeneous network topologies. For this, we create a hypergraph with $N = 20000$ nodes  and a power-law degree distribution of dyadic degrees with exponent $\gamma = 3.0$ and $k_{\text{max}}=N-1$ as described in Sec.~\ref{generativemodel}, again setting $\hat{q}_n = \hat{k}_n$. This allows us to probe the interplay between higher-order interactions and structural heterogeneity, specifically how heavy-tailed distributions influence the onset of synchronization and bistability.

The bottom panels of Fig.~\ref{fig:uniform_res} show the theoretical predictions against numerical simulation for both weak ($K_3 = 0.001485 < K_3^c = 0.002970$, left) and strong  ($K_3 = 0.007425 > K_3^c = 0.002970$, right) triadic coupling regimes. As in the uniform degree distribution case, the system exhibits a continuous transition for weak higher-order coupling and an explosive, discontinuous transition when $K_3$ exceeds the critical threshold. Overall, both the TAT and the FDA do a good job in this heterogeneous setting. 

%Unlike the FDA or standard mean-field theories (MFT) which often assume a thermodynamic limit or average over frequency realizations, the TAT approach accounts for the specific natural frequencies and degrees of individual nodes. Consequently, as seen in Fig.~\ref{fig:pw_res}, the TAT successfully captures the sample-specific fluctuations and finite-size effects that are characteristic of power-law networks, where highly-connected hubs can disproportionately influence the local order parameters. {\color

\subsection{Smaller Networks} \

Previous analyses of the Kuramoto model with higher-order interactions have relied on mean-field approaches valid for averaged networks or have been restricted to the all-to-all case in the $N\to \infty$ limit. These analyses rely on the self-averaging that occurs in large networks, where the particular frequencies of individual oscillators do not strongly influence the synchronization transition. The TAT presented in Sec.~\ref{sec:self-consistency}, on the other hand, takes into account the individual oscillator frequencies and hypergraph structure, making it especially suited to study the synchronization transition in relatively small ($N \sim 1000$) networks. 

\begin{figure}[b!]
    \centering

    \begin{subfigure}{0.48\textwidth}
        \centering
        \includegraphics[width=\linewidth]{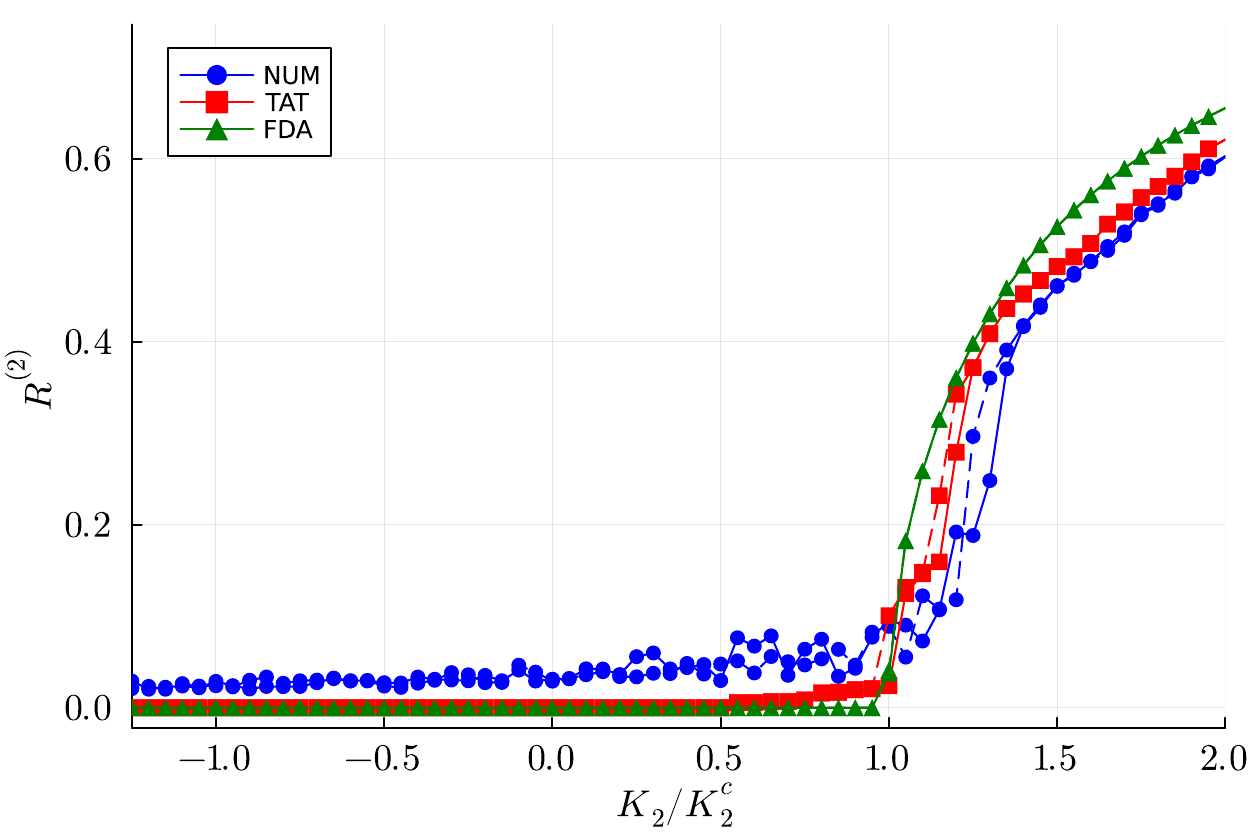}
        \caption{$K_3 < K_3^c$}
        \label{fig:small_k3lk3c}
    \end{subfigure}
    \hfill
    \begin{subfigure}{0.48\textwidth}
        \centering
        \includegraphics[width=\linewidth]{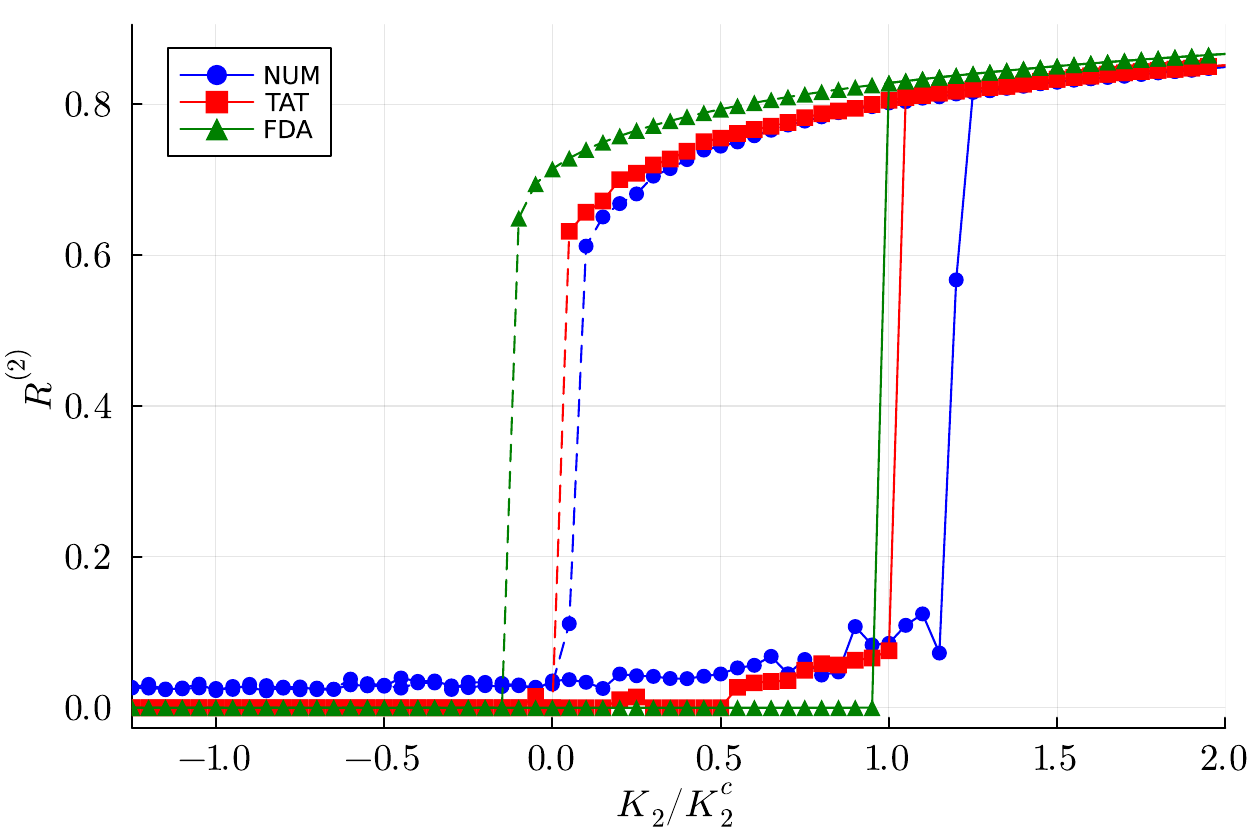}
        \caption{$K_3 > K_3^c$}
        \label{fig:small_k3gk3c}
    \end{subfigure}

    \caption{Examples of small network experiments with $N=1000,\langle k \rangle = \langle q \rangle=25, \gamma = 2.5$ with $k_{\text{max}}=125.$}
    \label{fig:small_nets}
\end{figure}

To illustrate the effectiveness of the TAT theory in this situation, in Fig.~\ref{fig:small_nets} we show the global order parameter $R^{(2)}$ obtained from direct numerical solution of Eq.~(\ref{eq:ode_sys}), from the TAT, and from the FDA for a hypergraph with $N = 1000$, and $\hat k_n = \hat q_n$ distributed as a power law with exponent $\gamma = 2.5$ and $k_{\text{max}}=125$. The left panel corresponds to $K_3 = 0.00832 < K_3^c = 0.0166$, while the right panel corresponds to $K_3 = 0.0416 > K_3^c = 0.0166$.  This case tests the limit of applicability of our approximations. For such a small, heterogeneous network, the presence of hubs and the particular frequencies they are assigned can significantly affect the transition to synchrony. For the specific realization shown in the figure, the FDA consistently predicts a smoother transition than what is  actually observed and large values of the order parameter in the synchronized regime. The TAT, on the other hand, captures the features of the synchronization transition much better.

\subsection{Onset of bistability} \label{sec:onset_bistability} \

A key feature distinguishing the Kuramoto model with higher-order interactions from the traditional dyadic Kuramoto model is the emergence of an abrupt (``explosive'') synchronization transition accompanied by bistability and hysteresis when the triadic coupling strength exceeds a critical threshold, $K_3 > K_3^c$. In Sec.~\ref{sec:PT_BI}, we derived analytical estimates for this threshold under the FDA framework, $K_3^c$ [Eq.~\eqref{eq:K3c_eq}], which incorporates the specific hypergraph topology via the Perron--Frobenius eigenvectors of $\boldsymbol{A}$ and the adjacency tensor $\boldsymbol{B}$, as well as under the HMF approximation, $K_3^{\text{HMF}}$ [Eq.~\eqref{newk3}], which depends only on moments of the degree distribution.

To test these predictions against direct numerical simulations of Eq.~\eqref{eq:ode_sys}, we quantify the extent of bistability as a function of $K_3$, which is varied in steps of $0.05 K_3^c$. For each value of $K_3$, we perform adiabatic forward and backward sweeps of the dyadic coupling strength $K_2$ over the interval $[0.5 K_2^c, 1.5 K_2^c]$, as described in Sec.~\ref{numerical}, and measure the maximum difference in the global order parameter between the two branches:
\begin{align}
    \Delta R^{(2)} = \max_{K_2 \in [0.5 K_2^c,\, 1.5 K_2^c]} \left| R^{(2)}_{\text{backward}}(K_2) - R^{(2)}_{\text{forward}}(K_2) \right|. \label{eq:delta_R2}
\end{align}
In the supercritical regime ($K_3 < K_3^c$), where the synchronization transition is continuous and reversible, $\Delta R^{(2)}$ remains close to zero up to finite-size fluctuations. Once $K_3$ surpasses the critical threshold ($K_3 > K_3^c$), the coexistence of coherent and incoherent states produces a hysteresis loop, resulting in a pronounced positive value of $\Delta R^{(2)}$. Below, we compare $\Delta R^{(2)}$ obtained from direct numerical simulations (NUM), the Time-Averaged Theory (TAT), the Frequency Distribution Approximation (FDA), and the Heterogeneous Mean-Field (HMF) approximation across hypergraphs with varying levels of degree heterogeneity.

\subsubsection{Homogeneous and moderately heterogeneous hypergraphs ($k_{\max} \ll N$)}\label{sssec:ceil_kmax} \

\begin{figure}[t!]
    \centering
    % --- First Row ---
    \begin{subfigure}{0.45\textwidth}
        \centering
        \includegraphics[width=\textwidth]{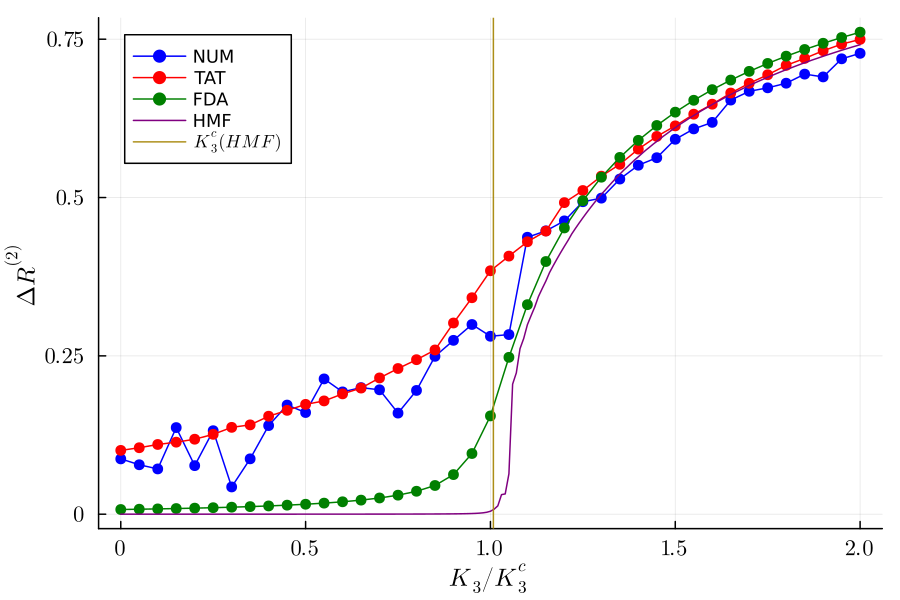}
        \caption{$\gamma = 2.5$}
        \label{fig:gamma25}
    \end{subfigure}
    \hfill
    \begin{subfigure}{0.45\textwidth}
        \centering
        \includegraphics[width=\textwidth]{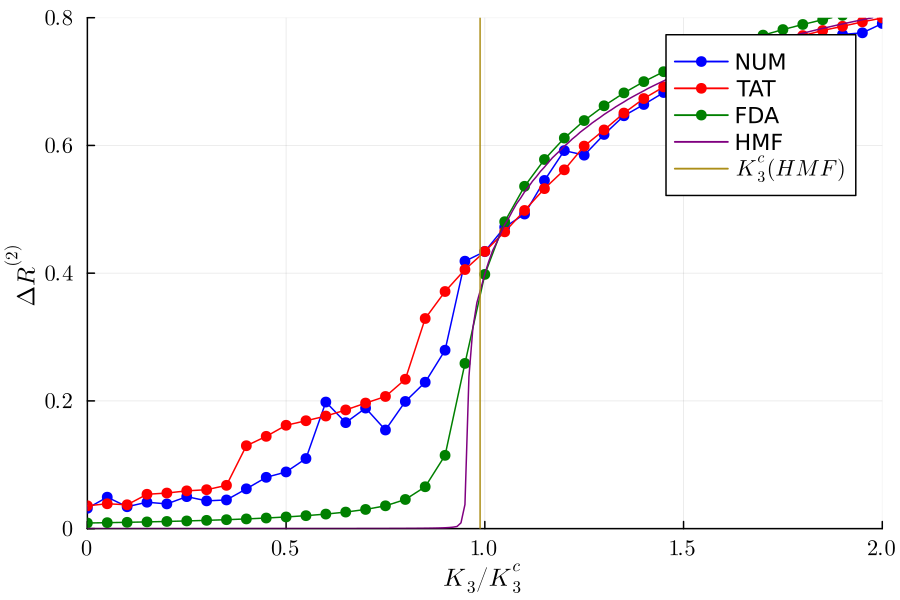}
        \caption{$\gamma = 3.0$}
        \label{fig:gamma30}
    \end{subfigure}

    \vspace{0.3cm}

    % --- Second Row ---
    \begin{subfigure}{0.45\textwidth}
        \centering
        \includegraphics[width=\textwidth]{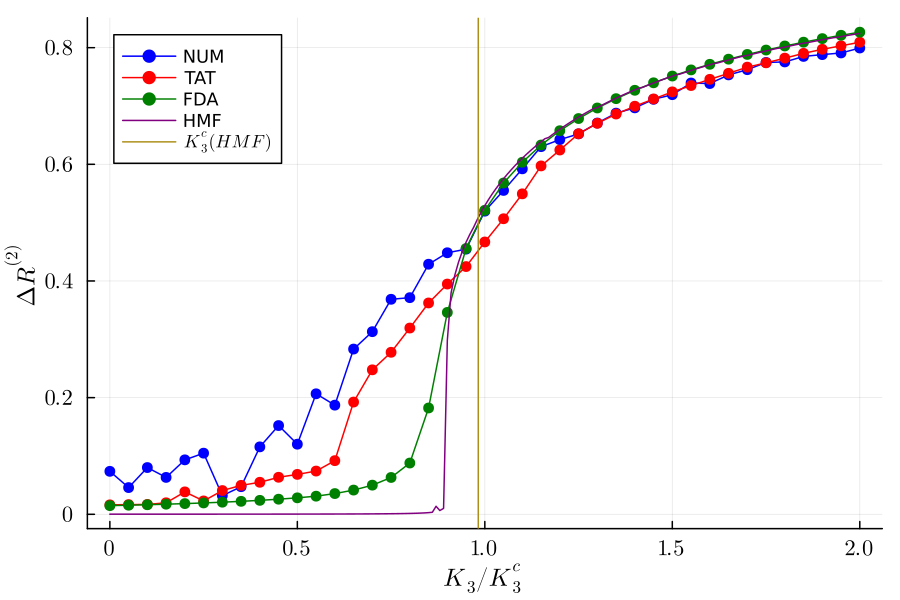}
        \caption{$\gamma = 4.0$}
        \label{fig:gamma40}
    \end{subfigure}
    \hfill
    \begin{subfigure}{0.45\textwidth}
        \centering
        \includegraphics[width=\textwidth]{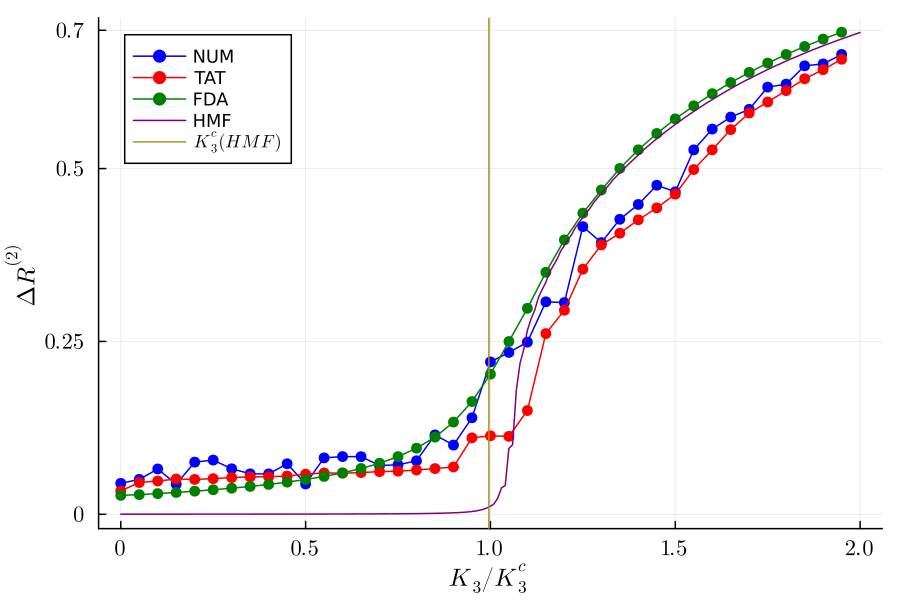}
        \caption{Uniform distribution}
        \label{fig:uni_dist}
    \end{subfigure}

    \caption{Maximum hysteresis difference $\Delta R^{(2)}$ as a function of $K_3/K_3^c$ for hypergraphs with $N = 5000$ and $\langle k \rangle = \langle q \rangle = 150$. Panels (a)--(c) correspond to power-law degree distributions with a capped maximum degree $k_{\max} = 1000$ for $\gamma = 2.5$, $3.0$, and $4.0$, respectively, while panel (d) shows a uniform degree distribution ($k_n, q_n \in [50, 250]$). The vertical gold line indicates the HMF prediction $K_3^{\text{HMF}}/K_3^c$ from Eq.~\eqref{newk3}.}
    \label{fig:bistab_capped}
\end{figure}

We first examine hypergraphs of size $N = 5000$ with mean degree $\langle k \rangle = \langle q \rangle = 150$ ($\hat{q}_n = \hat{k}_n$) where extreme hubs are absent. As a homogeneous benchmark, Fig.~\ref{fig:bistab_capped}(d) shows $\Delta R^{(2)}$ as a function of the normalized triadic coupling $K_3 / K_3^c$ for a uniform degree distribution. In this case, all four methods (NUM, TAT, FDA, and HMF) remain near zero for $K_3 < K_3^c$ and exhibit a sharp increase precisely at the predicted threshold $K_3 / K_3^c = 1$. Because degree fluctuations are small in a uniform distribution, the HMF threshold $K_3^{\text{HMF}}$ (indicated by the vertical gold line) is nearly identical to the spectral FDA prediction $K_3^c$.

Next, we consider power-law degree distributions with exponents $\gamma = 2.5$, $3.0$, and $4.0$, where the maximum degree is capped at $k_{\max} = 1000$ [Figs.~\ref{fig:bistab_capped}(a)--(c)]. Imposing the ceiling $k_{\max} \ll N$ prevents the formation of massive hubs and bounds the higher-order degree moments—most notably $\langle k^4 \rangle$ in Eq.~\eqref{oldk3}. Consequently, for $\gamma=3.0$ and $4.0$ the onset of bistability occurs near $K_3/K_3^c=1$ [Figs.~\ref{fig:bistab_capped}(b)--(c)], while for $\gamma=2.5$ the increase of $\Delta R^{(2)}$ is more gradual and begins below $K_3^c$ [Fig. \ref{fig:bistab_capped}(a)]. In all three cases, $K_3^{HMF}$ remains in close agreement with $K_3^c$.

Comparing the curves in Figs.~\ref{fig:bistab_capped}(a)--(c) also highlights the distinct advantages of the TAT in finite heterogeneous networks. While the FDA and HMF curves are smooth and remain nearly zero below $K_3^c$, the direct numerical simulations exhibit a non-zero baseline for $K_3 < K_3^c$ that grows as the network becomes more heterogeneous ($\gamma$ decreases from $4.0$ to $2.5$). Because the TAT retains the specific realization of natural frequencies and hypergraph connections, it accurately captures both this sub-threshold finite-size behavior and the growth of $\Delta R^{(2)}$ above $K_3^c$. Since the TAT contains no temporal fluctuations, the fact that it reproduces this baseline indicates that it reflects small, realization-specific hysteresis rather than noise; consistent with this, the baseline decreases as the hypergraph becomes more homogeneous [Fig.~\ref{fig:bistab_capped}(c)--(d)].

\subsubsection{Extreme degree heterogeneity ($k_{\max} = N$)}\label{sssec:uncapped_kmax} \

\begin{figure}[t!]
    \centering
    % --- First Row ---
    \begin{subfigure}{0.45\textwidth}
        \centering
        \includegraphics[width=\textwidth]{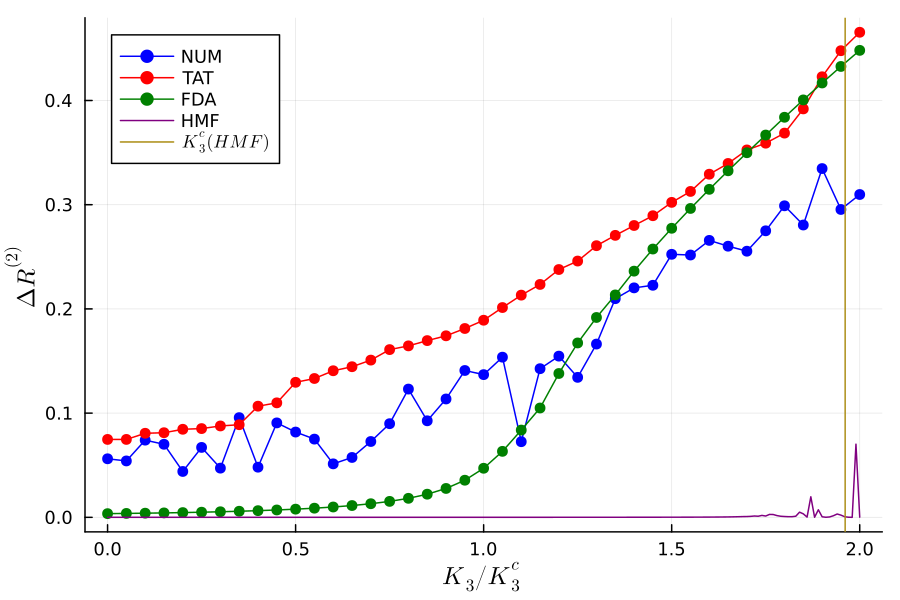}
        \caption{$\gamma = 2.5$}
        \label{fig:gamma25_kmax5k}
    \end{subfigure}
    \hfill
    \begin{subfigure}{0.45\textwidth}
        \centering
        \includegraphics[width=\textwidth]{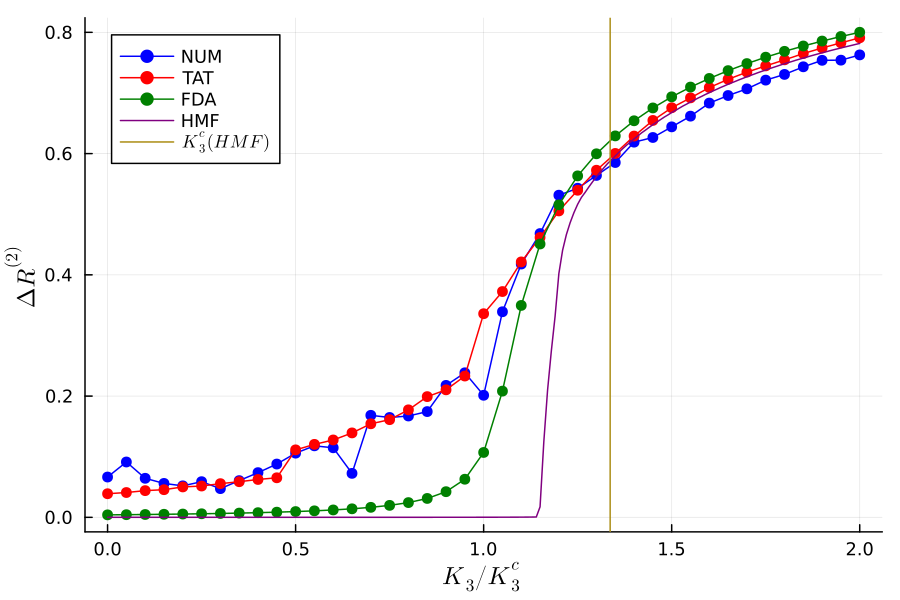}
        \caption{$\gamma = 3.0$}
        \label{fig:gamma30_kmax5k}
    \end{subfigure}

    \caption{Maximum hysteresis difference $\Delta R^{(2)}$ as a function of $K_3/K_3^c$ for power-law hypergraphs with an uncapped maximum degree $k_{\max} = N = 5000$ for (a) $\gamma = 2.5$ and (b) $\gamma = 3.0$. The vertical gold line denotes the HMF prediction $K_3^{\text{HMF}}/K_3^c$, which severely overestimates the threshold in the presence of extreme hubs.}
    \label{fig:bistab_uncapped}
\end{figure}

To test the limits of our theoretical approximations when degree heterogeneity becomes extreme, we relax the upper degree cutoff to the system size, $k_{\max} = N = 5000$, for power-law distributions with $\gamma = 2.5$ and $\gamma = 3.0$ (Fig.~\ref{fig:bistab_uncapped}). Allowing degrees up to $N$ permits the emergence of dominant hub nodes and causes the fourth moment $\langle k^4 \rangle$ (which diverges as $N \to \infty$ for $\gamma \le 5$) to become very large.

As shown in Fig.~\ref{fig:bistab_uncapped}, this extreme heterogeneity causes the HMF approximation to break down. For $\gamma = 3.0$ [Fig.~\ref{fig:gamma30_kmax5k}], Eq.~\eqref{oldk3} substantially overestimates the critical triadic coupling strength, predicting $K_3^{\text{HMF}} \approx 1.37 K_3^c$ (vertical gold line), while the numerical simulations, TAT, and FDA all show the onset of bistability near $K_3 / K_3^c = 1$. Thus, for $\gamma = 3.0$, incorporating the true spectral structure of the hypergraph via the FDA and TAT successfully corrects the failure of the degree-based HMF theory.

When the heterogeneity is increased further to $\gamma = 2.5$ with $k_{\max} = 5000$ [Fig.~\ref{fig:gamma25_kmax5k}], the HMF threshold shifts even further to $K_3^{\text{HMF}} \approx 1.96 K_3^c$, and the HMF curve fails to predict any bistability across nearly the entire plotted range. Moreover, in this extreme regime, noticeable discrepancies also emerge between the self-consistent theories (TAT and FDA) and the direct numerical simulations: the sharp onset of bistability is smoothed out, and both TAT and FDA overestimate $\Delta R^{(2)}$ for $K_3 > K_3^c$. This deviation reflects the breakdown of the simplifying assumptions summarized in Sec.~\ref{assumptions} when a few massive hubs dominate the hypergraph. Specifically, strong hub nodes disproportionately drive their neighbors, inducing correlations between local mean fields $H_n$ and natural frequencies $\omega_n$ (violating Assumption 3) as well as dynamical pair correlations (violating Assumption 5). Together, Figs.~\ref{fig:bistab_capped} and \ref{fig:bistab_uncapped} confirm that while the TAT and FDA accurately capture the synchronization transition and onset of bistability for specific hypergraph realizations across a wide range of topologies, their quantitative accuracy relies on the network not being excessively dominated by extreme hubs.

\subsection{Effect of hypergraph correlations on the onset of bistability} \label{sec:topo_graph} \

In this section, we explore how the critical triadic coupling strength $K_3^c$ governing the onset of bistability depends on structural correlations between the dyadic and triadic interactions. Following the analytical results derived in Sec.~\ref{sec:PT_BI}, we first conduct numerical experiments to examine how correlations between the nodal dyadic and triadic degrees shift the onset of bistability. Next, we demonstrate how hypergraphs can be rewired at fixed degree distributions to induce or suppress bistability by modifying eigenvector-tensor correlations.

\subsubsection{Dyadic and triadic degree correlations} \label{sssec:deg_corr} \

As shown in Sec.~\ref{sec:PT_BI}, the HMF approximation [Eqs.~\eqref{newk3} and \eqref{newk3unco}] predicts that the critical triadic coupling strength $K_3^{\text{HMF}}$ depends directly on correlations between the dyadic degrees $k_n$ and triadic degrees $q_n$ via the joint moments $\langle q k \rangle$ and $\langle q k^2 \rangle$. Specifically, $K_3^c$ is predicted to be smallest when dyadic and triadic degrees are positively correlated (maximizing the denominator of Eq.~\eqref{newk3}) and largest when they are anti-correlated.

To test this prediction, we construct three hypergraphs of size $N = 5000$ whose dyadic and triadic degrees are positively correlated, uncorrelated, and anti-correlated, respectively. First, we sample a sequence of target dyadic degrees $\{\hat{k}_n\}_{n=1}^N$ from a uniform distribution on $[50, 250]$ (yielding $\langle \hat{k} \rangle = 125$). We then assign the target triadic degrees $\{\hat{q}_n\}_{n=1}^N$ as follows:
\begin{enumerate}
    \item \textit{Correlated case:} We set the target triadic degrees equal to the target dyadic degrees, $\hat{q}_n = \hat{k}_n$.
    \item \textit{Uncorrelated case:} We generate a random permutation $\pi$ of $\{1, 2, \dots, N\}$ and set $\hat{q}_n = \hat{k}_{\pi(n)}$.
    \item \textit{Anti-correlated case:} After sorting the target dyadic degrees in ascending order, $\hat{k}_1 \leq \hat{k}_2 \leq \dots \leq \hat{k}_N$, we set $\hat{q}_n = \hat{k}_{N-n+1}$.
\end{enumerate}
Using these target degree sequences, we generate the corresponding hypergraphs according to the Chung--Lu procedure described in Sec.~\ref{generativemodel}. To quantify the emergence of bistability across these three topologies, we compute the maximum hysteresis difference $\max |\Delta R| \equiv \Delta R^{(2)}$ between the forward and backward $K_2$ sweeps [Eq.~\eqref{eq:delta_R2}] as a function of $K_3 / K_3^{c,\text{corr}}$, where $K_3^{c,\text{corr}}$ is the critical triadic coupling strength calculated from Eq.~\eqref{eq:K3c_eq} for the correlated hypergraph.

\begin{figure}[t!]
    \centering
    % --- First Row ---
    \begin{subfigure}{0.48\textwidth}
        \centering
        \includegraphics[width=\textwidth]{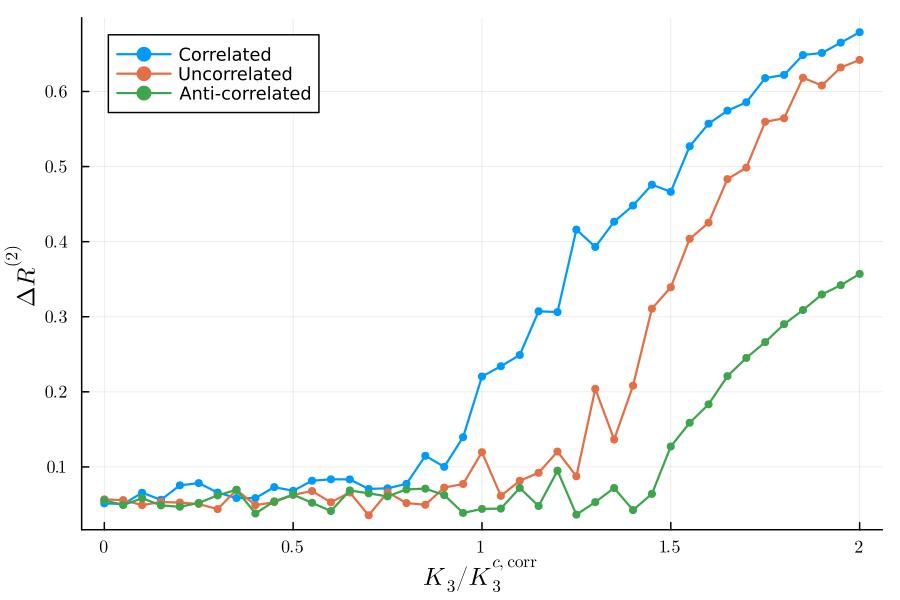}
        \caption{Numerical results (NUM)}
        \label{fig:NUM_corr}
    \end{subfigure}
    \hfill
    \begin{subfigure}{0.48\textwidth}
        \centering
        \includegraphics[width=\textwidth]{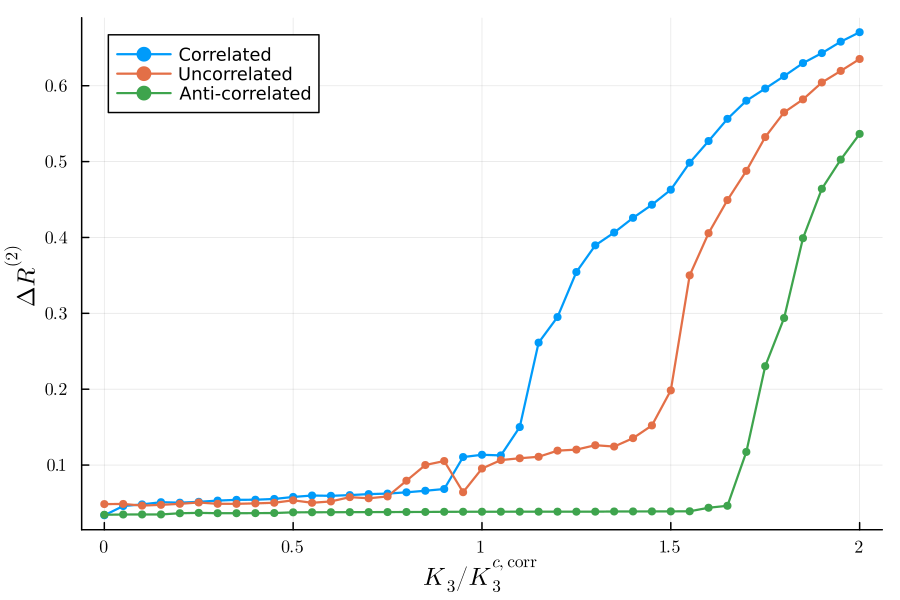}
        \caption{TAT results}
        \label{fig:TAT_corr}
    \end{subfigure}

    \vspace{0.3cm}

    % --- Second Row ---
    \begin{subfigure}{0.48\textwidth}
        \centering
        \includegraphics[width=\textwidth]{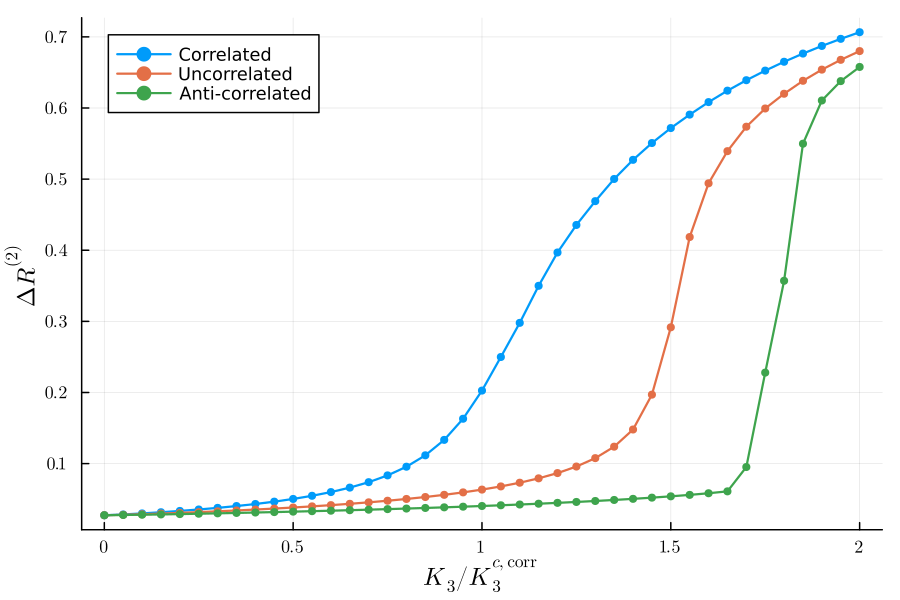}
        \caption{FDA results}
        \label{fig:FDA_corr}
    \end{subfigure}
    \hfill
    \begin{subfigure}{0.48\textwidth}
        \centering
        \includegraphics[width=\textwidth]{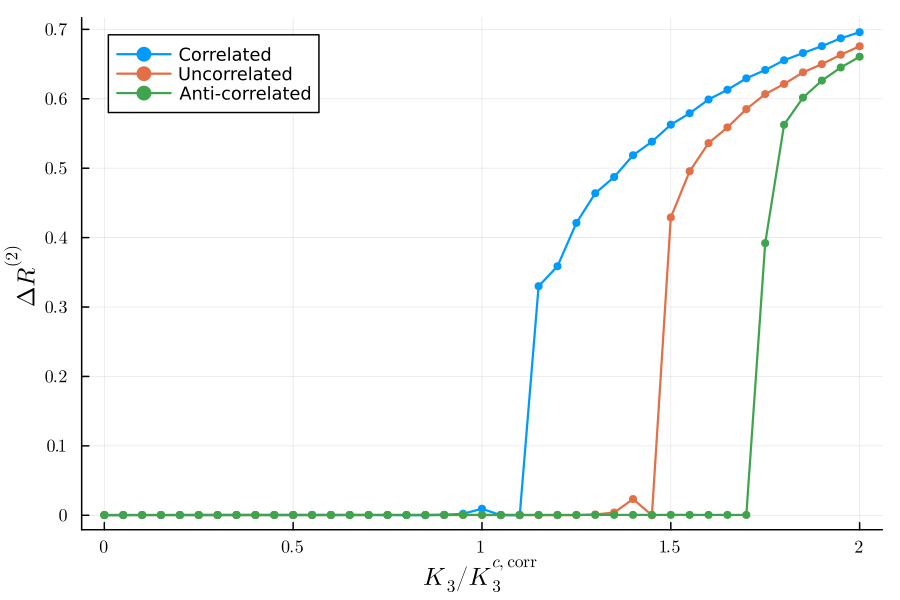}
        \caption{HMF results}
        \label{fig:HMF_corr}
    \end{subfigure}
    \caption{Maximum hysteresis difference $\max |\Delta R|$ versus $K_3/K_3^{c,\text{corr}}$ for hypergraphs with correlated (blue circles), uncorrelated (orange circles), and anti-correlated (green circles) dyadic and triadic degrees ($N = 5000$, uniform degree distribution in $[50, 250]$). The four panels compare (a) direct numerical simulations of Eq.~\eqref{eq:ode_sys}, (b) TAT, (c) FDA, and (d) HMF. In all panels, $K_3$ is normalized by the critical value $K_3^{c,\text{corr}}$ calculated from Eq.~\eqref{eq:K3c_eq} for the correlated hypergraph.}
    \label{fig:compare_corr}
\end{figure}

Figure~\ref{fig:compare_corr} compares $\max |\Delta R|$ across all four methods—direct numerical simulations [Fig.~\ref{fig:NUM_corr}], TAT [Fig.~\ref{fig:TAT_corr}], FDA [Fig.~\ref{fig:FDA_corr}], and HMF [Fig.~\ref{fig:HMF_corr}]—for the correlated (blue circles), uncorrelated (orange circles), and anti-correlated (green circles) hypergraphs. Across all four panels, the relative ordering predicted by Eqs.~\eqref{eq:K3c_eq} and \eqref{newk3} holds: the correlated hypergraph undergoes the onset of bistability at the smallest value of $K_3$ ($K_3 / K_3^{c,\text{corr}} \approx 1$), followed by the uncorrelated hypergraph ($K_3 / K_3^{c,\text{corr}} \approx 1.4$), while the anti-correlated hypergraph requires the largest triadic coupling strength ($K_3 / K_3^{c,\text{corr}} \approx 1.6$) to exhibit bistability.

Comparing the theoretical approximations against the direct numerical simulations in Fig.~\ref{fig:NUM_corr} also highlights the distinct characteristics of each approximation level. While the HMF [Fig.~\ref{fig:HMF_corr}] and FDA [Fig.~\ref{fig:FDA_corr}] accurately capture the directional trend and relative spacing of the critical thresholds across the three correlation regimes, both average over the oscillator frequency distribution and therefore miss the finite-size fluctuations of the realized system. Specifically, HMF predicts an abrupt jump from zero at each threshold, and both FDA and HMF overestimate the magnitude of the hysteresis loop in the anti-correlated case at large $K_3$ ($\max |\Delta R| \approx 0.66$ at $K_3 / K_3^{c,\text{corr}} = 2$, compared to $\approx 0.36$ in the numerical simulations). In contrast, because the TAT [Fig.~\ref{fig:TAT_corr}] incorporates both the specific hypergraph structure and the realization of natural frequencies, it captures the non-zero pre-transitional baseline ($\max |\Delta R| \approx 0.05$) as well as the more gradual growth of the hysteresis amplitude observed in the numerical simulations.

\subsubsection{Rewiring hypergraphs} \label{rewiring} \

Equation (\ref{eq:K3c_eq}) shows that the critical triadic coupling strength, $K_3^c$, depends on how dyadic and triadic interactions are correlated. Focusing on the denominator of Eq.~(\ref{eq:K3c_eq}), $\sum_{n,m,j} \boldsymbol{B}_{nmj}  V_n U_m^2 U_j$, we note that an increase in its value lowers $K_3^c$, and vice versa. This suggests that increasing (decreasing) the value of the product $V_n U_m^2 U_j$ over hyperedges $\{n,m,j\}$ by rewiring hyperedges should decrease (increase) $K_3^c$, provided the assumptions under which Eq.~(\ref{eq:K3c_eq}) was derived hold.

\begin{figure}[t!]
    \centering
    % --- First Row ---
    \begin{subfigure}{0.45\textwidth}
        \centering
        \includegraphics[width=\textwidth]{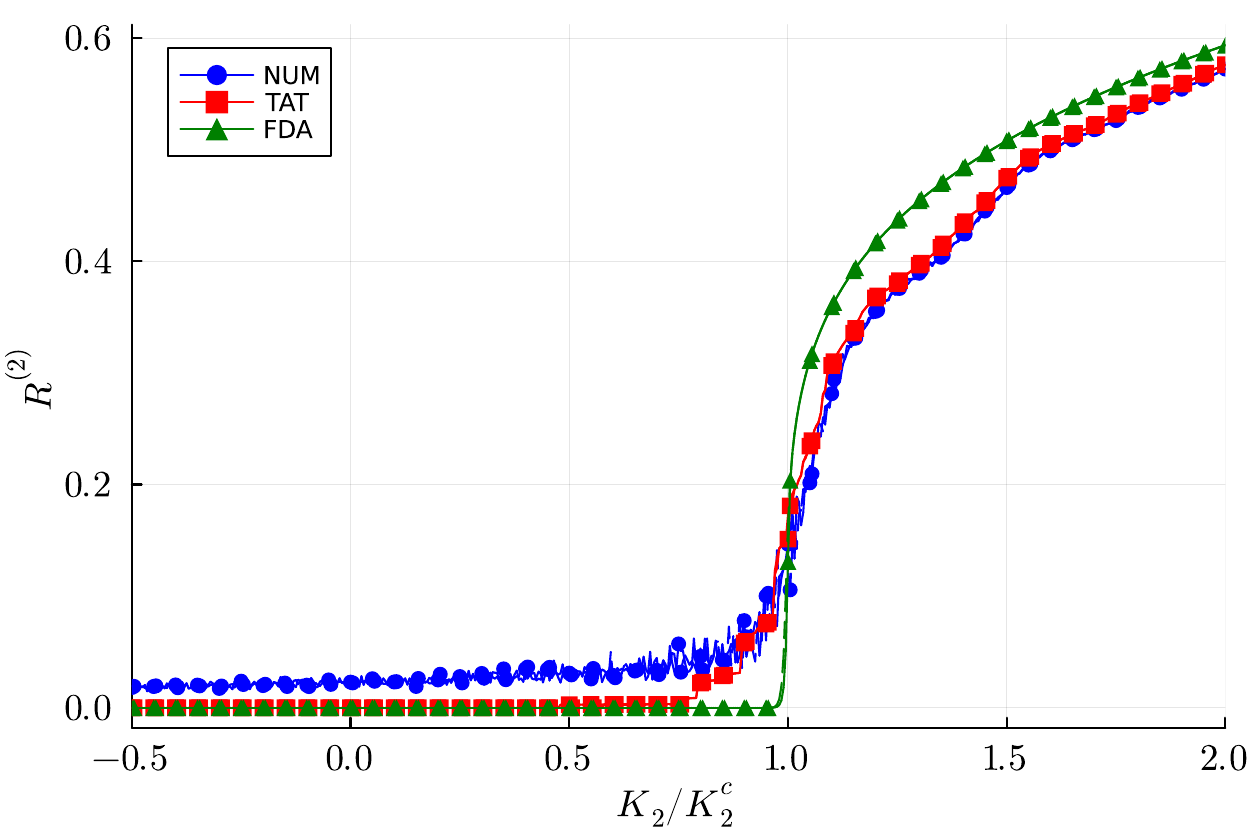}
        \caption{Original}
        \label{fig:orig_de}
    \end{subfigure}
    \hfill
    \begin{subfigure}{0.45\textwidth}
        \centering
        \includegraphics[width=\textwidth]{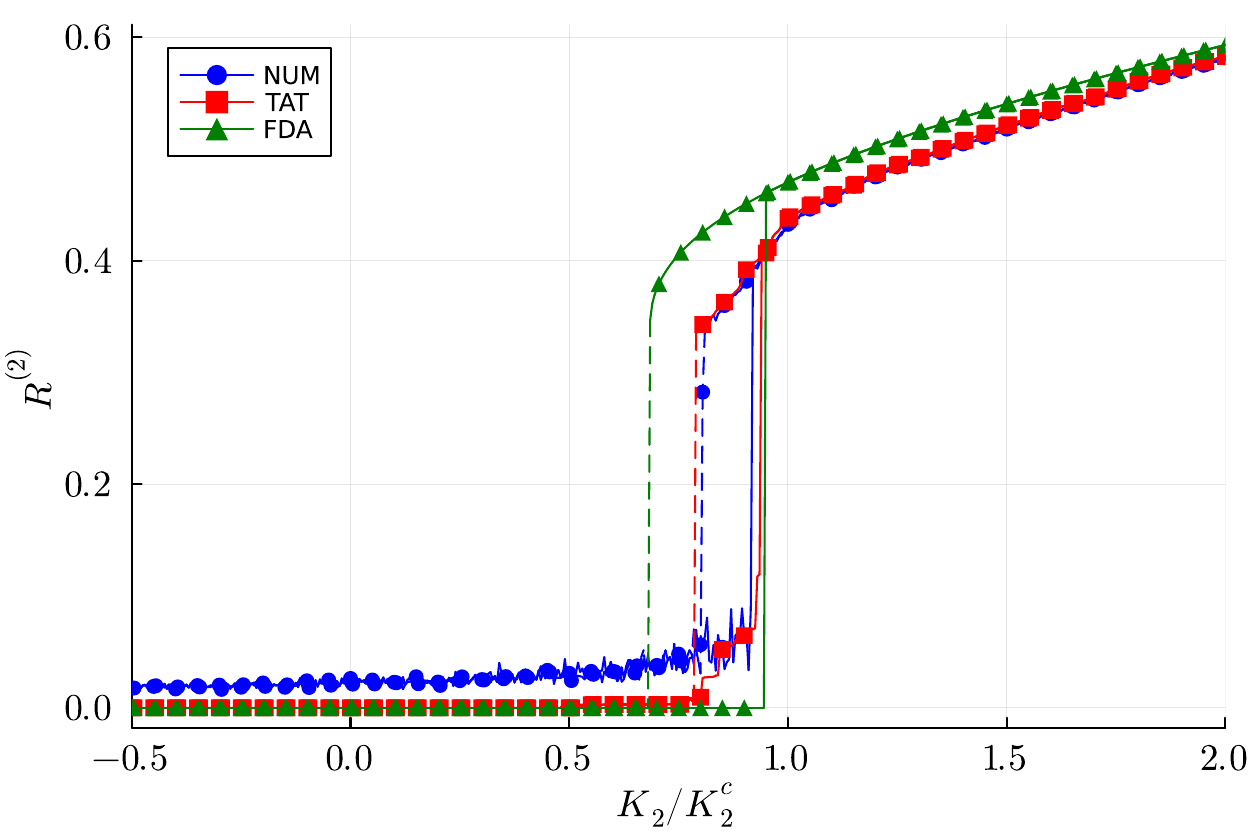}
        \caption{Rewired to decrease $K^c_3$}
        \label{fig:rewire_de}
    \end{subfigure}

    \vspace{0.3cm} % Adds a bit of vertical spacing between rows

    % --- Second Row ---
    \begin{subfigure}{0.45\textwidth}
        \centering
        \includegraphics[width=\textwidth]{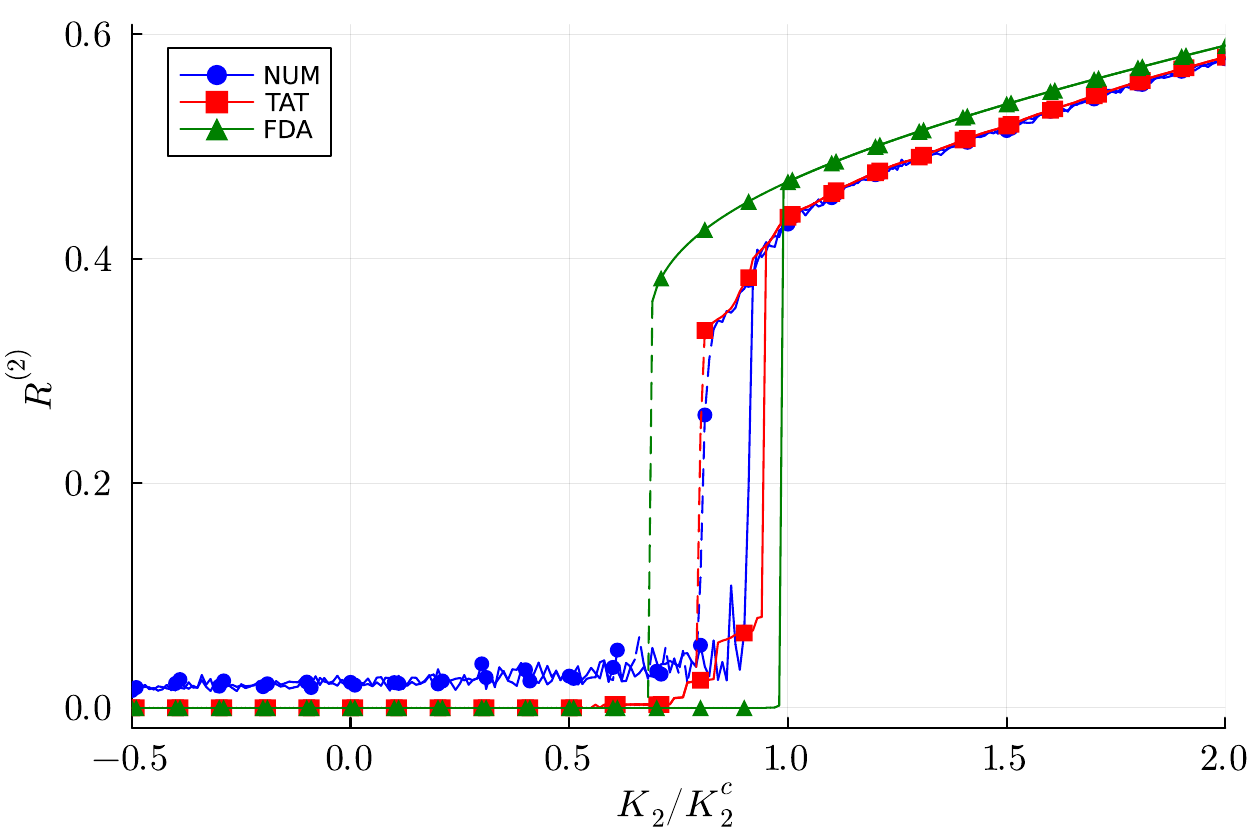}
        \caption{Original}
        \label{fig:orig_in}
    \end{subfigure}
    \hfill
    \begin{subfigure}{0.45\textwidth}
        \centering
        \includegraphics[width=\textwidth]{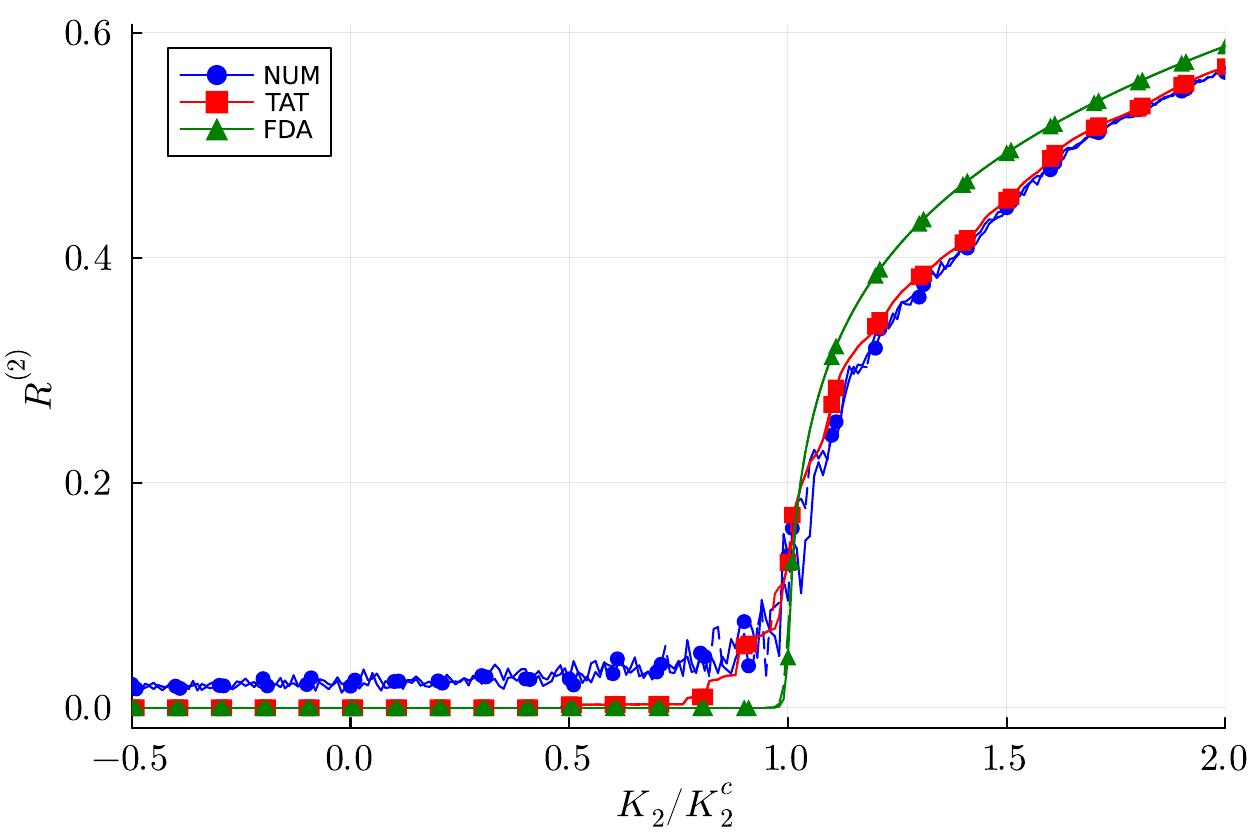}
        \caption{Rewired to increase $K^c_3$}
        \label{fig:rewire_in}
    \end{subfigure}

    \caption{Comparison of the synchronization transition, $R^{(2)}$ versus $K_2/K_2^c$, for original (left) and rewired (right) hypergraphs, demonstrating structural control over the onset of bistability. Panels (a) and (b) compare the system at the same triadic coupling strength $K_3$, where (a) shows the original hypergraph ($K_3 < K_3^{\text{orig}}$) and (b) shows the rewired hypergraph with a lowered critical coupling ($K_3^{\text{low}} < K_3 < K_3^{\text{orig}}$). Conversely, panels (c) and (d) compare the system at a higher fixed $K_3$, where (c) shows the original hypergraph ($K_3 > K_3^{\text{orig}}$) and (d) shows the rewired hypergraph with an increased critical coupling ($K_3^{\text{orig}} < K_3 < K_3^{\text{high}}$). The hypergraph is constructed with $N=4000$, $\langle k \rangle = \langle q \rangle = 100$, $k_{\min}=50$, and $k_{\max}=400$.}
    \label{fig:compare_rewiring}
\end{figure}

To test this prediction, we first construct a hypergraph with a bimodal distribution of target degrees,
\begin{align}
\hat k_n = \hat q_n =
\begin{cases} 
k_{\max}, & \text{with probability } \frac{\langle k \rangle - k_{\min}}{k_{\max} - k_{\min}}, \\
k_{\min}, & \text{otherwise},
\end{cases}
\label{eq:bimodal_dist}
\end{align}
where we set $N=4000$, $k_{\min}=50$, $k_{\max}=400$, and $\langle k \rangle = 100$. With these parameters, approximately $14\%$ of the nodes are assigned a degree $\hat{k}_n=400$. Using the procedure described in Sec.~\ref{generativemodel}, we construct the {\it original} hypergraph characterized by the adjacency matrix and tensor $\boldsymbol{A}^{\text{orig}}$, $\boldsymbol{B}^{\text{orig}}$. We denote the critical triadic coupling strength for this hypergraph, calculated from (\ref{eq:K3c_eq}), as $K_3^{\text{orig}}$. Then, following the procedure described in Appendix~\ref{appx:rewire_alg}, we rewire the triadic edges resulting in a new adjacency tensor $\boldsymbol{B}^{\text{low}}$ with a larger value of the sum $\sum_{n,m,j} \boldsymbol{B}_{nmj}  V_n U_m^2 U_j$. Therefore, this rewired hypergraph results in a lower value of the critical triadic coupling strength, $K_3^{\text{low}} < K_3^{\text{orig}}$. In order to compare the original and rewired hypergraphs, we simulate Eqs.~(\ref{eq:ode_sys}) using a value of $K_3$ satisfying $K_3^{\text{low}} < K_3 < K_3^{\text{orig}}$. In Figure~\ref{fig:compare_rewiring}(a) we show the order parameter as a function of $K_2$ for the original hypergraph, for which $K_3 < K_3^{\text{orig}}$. As expected, there is no bistability and the transition to synchrony appears to be supercritical. On the other hand, for the rewired hypergraph, for which the same quantities are shown in Fig.~\ref{fig:compare_rewiring}(b) and for which $K_3^{\text{low}} < K_3$, there is a bistable regime. Thus, the rewiring algorithm is able to successfully induce bistability in the dynamics by modifying the correlations between dyadic and triadic structure.

Interestingly, rewiring designed to increase the predicted critical value $K_3^c$ was not always effective when applied directly to the original hypergraph. In contrast, a two-step procedure consisting of first rewiring to decrease $K_3^c$, and then rewiring in the opposite direction to increase $K_3^c$, more reliably eliminated the bistable region. This suggests that structural control of bistability is path-dependent and that the scalar quantity appearing in Eq.~(\ref{eq:K3c_eq}) does not fully characterize the hypergraph features controlling the nonlinear bistable regime. In particular, higher-order correlations and finite-size effects may influence the outcome of the rewiring procedure beyond what is predicted by the perturbative analysis. With this caveat, Figures~\ref{fig:compare_rewiring}(c) and (d) show the results of a similar experiment designed to increase the critical triadic coupling strength. In this case, we again start with an original hypergraph as above. Then, we first rewire the network to increase the sum $\sum_{n,m,j} \boldsymbol{B}_{nmj}  V_n U_m^2 U_j$, and then rewire it to decrease it, finally resulting in a hypergraph with an adjacency tensor $\boldsymbol{B}^{\text{high}}$ with a smaller value of the sum $\sum_{n,m,j} \boldsymbol{B}_{nmj}^{\text{high}}  V_n U_m^2 U_j$ than the original hypergraph. The final hypergraph results, then, in a higher critical triadic coupling strength $K_3^{\text{high}} > K_3^{\text{orig}}$. As in the previous experiment, we then simulate Eqs.~(\ref{eq:ode_sys}) using a value of $K_3$ satisfying $K_3^{\text{orig}} < K_3 < K_3^{\text{high}}$. Figure~\ref{fig:compare_rewiring}(c) shows that for the original hypergraph, satisfying $K_3^{\text{orig}} < K_3$, there is bistability, while for the rewired hypergraph, which satisfies $K_3 < K_3^{\text{high}}$, the bistable region is eliminated. 

Note, by the way, how the TAT reproduces extremely well the features of the synchronization transition in all the panels of Fig.~\ref{fig:compare_rewiring}, in contrast with the FDA.

\section{Discussion and Conclusion}
\label{sec:discussion}

In this work, we developed a self-consistent analytical framework for the Kuramoto model with dyadic and triadic interactions on hypergraphs. By introducing generalized local order parameters, we derived a three-tier hierarchy of approximations that balances analytical tractability and required system information against predictive accuracy:

\begin{itemize}
    \item \textbf{Time-Averaged Theory (TAT):} By retaining both the exact hypergraph topology ($\boldsymbol{A}$ and $\boldsymbol{B}$) and the specific realization of intrinsic frequencies $\{\omega_n\}$, the TAT captures sample-specific finite-size fluctuations, pre-transitional baselines, and local synchronization effects that are smoothed out by ensemble averages. Although it requires complete microscopic knowledge of the system, the TAT provides superior accuracy in relatively small networks ($N \sim 1000$) and heterogeneous topologies (see Figs.~\ref{fig:small_nets}, \ref{fig:bistab_capped}, and \ref{fig:compare_rewiring}).
    
    \item \textbf{Frequency Distribution Approximation (FDA):} Averaging over the oscillator frequency distribution while preserving the exact hypergraph structure eliminates the need for individual frequency data. While the FDA cannot capture fluctuations driven by specific frequency assignments, it retains the spectral properties of the network. This enables closed-form perturbative analyses of the synchronization and bistability thresholds [Eqs.~\eqref{eq:eig_val_A} and \eqref{eq:K3c_eq}] that remain accurate across moderately heterogeneous hypergraphs (Figs.~\ref{fig:bistab_capped} and \ref{fig:compare_corr}).
    
    \item \textbf{Heterogeneous Mean-Field (HMF) Approximation:} Averaging over both frequency and hypergraph realizations requires only the degree sequences (or distributions) of the system, recovering previous results derived via the Ott--Antonsen ansatz~\cite{Adhikari2023}. While computationally simple, the HMF approximation is the least accurate in our hierarchy; it fails to predict the onset of bistability in hypergraphs with pronounced degree heterogeneity, where diverging higher-order degree moments such as $\langle k^4 \rangle$ skew the mean-field threshold (Fig.~\ref{fig:bistab_uncapped}).
\end{itemize}

Taken together, our results highlight that accurately predicting synchronization on hypergraphs requires accounting for both structural and dynamical realizations. The appropriate theoretical framework depends on both the available system information and the degree of structural heterogeneity. While HMF is sufficient for large, homogeneous networks, the FDA and TAT become essential as degree heterogeneity increases or system size decreases. Notably, unlike in the purely dyadic Kuramoto model, drifting oscillators make a non-vanishing contribution to the triadic local order parameter $R_n^{(3)}$ [Eq.~\eqref{eq:r3_tat_system}], playing a direct role in sustaining the bistable regime.

A central analytical contribution of this paper is the perturbative expression for the critical triadic coupling strength $K_3^c$ governing the onset of bistability and explosive synchronization [Eq.~\eqref{eq:K3c_eq}]. Derived within the FDA framework, Eq.~\eqref{eq:K3c_eq} reveals that the emergence of hysteresis is governed by the alignment between the Perron--Frobenius eigenvectors ($\boldsymbol{U}, \boldsymbol{V}$) of the dyadic adjacency matrix $\boldsymbol{A}$ and the triadic adjacency tensor $\boldsymbol{B}$. Specifically, bistability is promoted ($K_3^c$ is lowered) when triadic hyperedges preferentially couple nodes with large dyadic eigenvector centralities [Fig.~\ref{fig:corr_swap}(a)--(b)], or when nodal dyadic and triadic degrees are positively correlated [Eq.~\eqref{newk3}, Fig.~\ref{fig:corr_swap}(c)--(d), and Fig.~\ref{fig:compare_corr}]. Leveraging this spectral relationship, we demonstrated that targeted hyperedge rewiring at a fixed degree distribution can shift $K_3^c$ to either induce or suppress explosive synchronization and hysteresis (Fig.~\ref{fig:compare_rewiring}). However, the path-dependence observed when rewiring to suppress bistability indicates that the perturbative threshold $K_3^c$ does not fully govern the deeply nonlinear bistable regime, where higher-order structural correlations and finite-size effects become significant.

Our analytical framework relies on several simplifying assumptions (summarized in Sec.~\ref{assumptions}) that hold best for large, dense, and moderately heterogeneous hypergraphs. As demonstrated in Fig.~\ref{fig:bistab_uncapped}(a) for uncapped power-law networks with $\gamma = 2.5$, the presence of extreme hubs can violate the assumed statistical independence between local mean fields $H_n$ and natural frequencies $\omega_n$, as well as the neglect of dynamical pair correlations. Establishing rigorous error bounds for these approximations, or extending the self-consistent equations to incorporate pair correlations—analogous to pair-approximation methods in hypergraph contagion models~\cite{matamalas2020abrupt}—represents a promising direction for future work.

Finally, the methodology developed here can be naturally extended to broader classes of oscillator systems. Relaxing the single-cluster assumption ($\psi_n = 0$) by allowing node- or community-dependent local mean-field phases $\psi_n$ would enable the study of chimera states, traveling waves, and multi-cluster synchronization in modular or spatially embedded hypergraphs. Moreover, our self-consistent analysis could be adapted to alternative higher-order phase-coupling functions (for which the resulting expressions for $R^{(3)}_n$ and $K_3^c$, and therefore the conclusions about bistability, would differ), such as those containing harmonics like $\sin(\theta_m + \theta_j - 2\theta_n)$ that emerge alongside Eq.~\eqref{eq:ode_sys} in higher-order phase reductions~\cite{Leon2019}.

\ack{The authors gratefully acknowledge useful discussions with Sabina Adhikari and Per Sebastian Skardal.}

\funding{CK acknowledges partial funding support by the Interdisciplinary Quantitative Biology (IQ Biology) PhD program at the BioFrontiers Institute, University of Colorado at Boulder, and the National Science Foundation NRT Integrated Data Science Fellowship [award
2022138].}
% This section is a list of funder names and grant numbers

\data{Code is available at \cite{KumpeerakijGitHub}}.
% For more information on IOP Publishing's research data policy see: https://publishingsupport.iopscience.iop.org/questions/research-data/

\bibliographystyle{iopart-num}
\bibliography{ref}

\section*{Appendix}
\appendix

\section{Derivation of Eqs.~\eqref{eq:reduce_dyn_eq}--\eqref{eq:hn}} \label{appx:reduce_dyn}

Using the definitions in Eqs.~\eqref{eq:local_R2_def}--\eqref{eq:local_R3_def}, 
\begin{align}
    \sum_m A_{nm} \sin(\theta_m-\theta_n) &= \text{Im} [e^{-i\theta_n}\sum_m A_{nm} e^{i\theta_m}] = R^{(2)}_n\sin(\psi^{(2)}_n-\theta_n), \\
    \sum_{m,j} B_{nmj}\sin(2\theta_m-\theta_j-\theta_n) &= \text{Im}[e^{-i\theta_n} \sum_{m,j} B_{nmj}e^{i(2\theta_m-\theta_j)}]=R^{(3)}_n\sin(\psi^{(3)}_n-\theta_n). 
\end{align}

Therefore Eq.~\eqref{eq:ode_sys} becomes
\begin{align}
    \dot{\theta}_n &= \omega_n + K_2R^{(2)}_n\sin(\psi^{(2)}_n-\theta_n)+K_3R^{(3)}_n\sin(\psi^{(3)}_n-\theta_n)=\omega_n + \text{Im}[e^{-i\theta_n}H_ne^{i\psi_n}],
\end{align}
with $H_ne^{i\psi_n}=K_2R^{(2)}_ne^{i\psi^{(2)}_n}+K_3R^{(3)}_ne^{i\psi^{(3)}_n}$ [Eq.~\eqref{eq:hn}], and $\text{Im}[e^{-i\theta_n}H_ne^{i\psi_n}]=H_n\sin(\psi_n-\theta_n)=-H_n\sin(\theta_n-\psi_n)$, which gives Eq.~\eqref{eq:reduce_dyn_eq}.

\section{Derivation of Eq.~(\ref{eq:r2_tat})} \label{sec:real-term-appx} \

To derive Eq.~(\ref{eq:r2_tat}), we need to show that the second and third terms on the right-hand side of Eq.~(\ref{eq:each_term_R2}) are zero. The second term can be rewritten as
\begin{align}
    \sum_{|\omega_m| \le H_m} \boldsymbol{A}_{nm} \left(\frac{\omega_m}{H_m}\right) &= \sum_{m=1}^N \boldsymbol{A}_{nm} \left(\frac{\omega_m}{H_m}\right) I_{[-H_m,H_m]}(\omega_m) \\
    &= \left(\sum_{m=1}^N \boldsymbol{A}_{nm}\right)E_n\left[\left(\frac{\omega}{H}\right)I_{[-H,H]}(\omega)\right],
\end{align}
where $I_B(\omega)$ is the indicator function over set $B$ and $E_n[x] \equiv \sum_m \boldsymbol{A}_{nm} x_m / \sum_m \boldsymbol{A}_{nm}$ is an average over nodes connected to node $n$.
If the number of nodes connected to node $n$ is large and the values of $H_m$ and $\omega_m$ for these nodes are representative of the overall distribution, we can approximate 
\begin{align}
E_n\left[\left(\frac{\omega}{H}\right)I_{[-H,H]}(\omega)\right] \approx \int_0^{\infty} \int_{-\infty}^{\infty} g(\omega)h(H)\left(\frac{\omega}{H}\right) I_{[-H,H]}(\omega) d\omega dH,
\end{align}
where we used the assumption that $\omega$ and $H$ are statistically independent to write their joint distribution as a product of their marginal distributions $g(\omega)$ and $h(H)$. Using our assumption that the frequency distribution $g(\omega)$ is symmetric, we get
\begin{align}
E_n\left[\left(\frac{\omega}{H}\right)I_{[-H,H]}(\omega)\right] &\approx \int_0^{\infty} h(H)\left(\frac{1}{H}\right)\left(\int_{-\infty}^{\infty} g(\omega) \omega I_{[-H,H]}(\omega) d\omega \right)dH = 0,
\end{align}
since the integral in parentheses vanishes by symmetry.

\begin{figure}[t!]
    \centering
    \begin{subfigure}{0.48\textwidth}
        \centering
        \includegraphics[width=\linewidth]{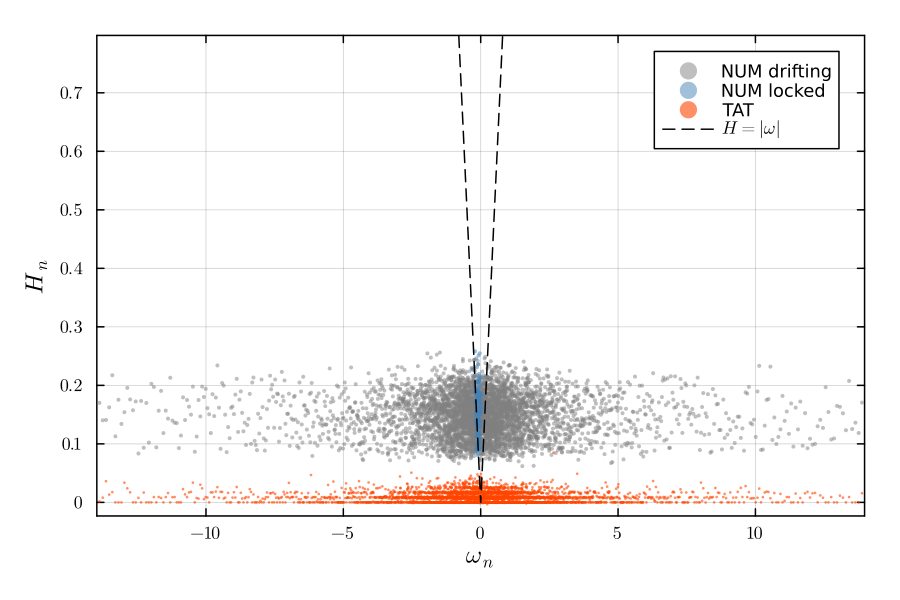}
        \caption{$K_2 = 0.8 K_2^c$}
        \label{fig:Hn_omega_k2_08}
    \end{subfigure}
    \hfill
    \begin{subfigure}{0.48\textwidth}
        \centering
        \includegraphics[width=\linewidth]{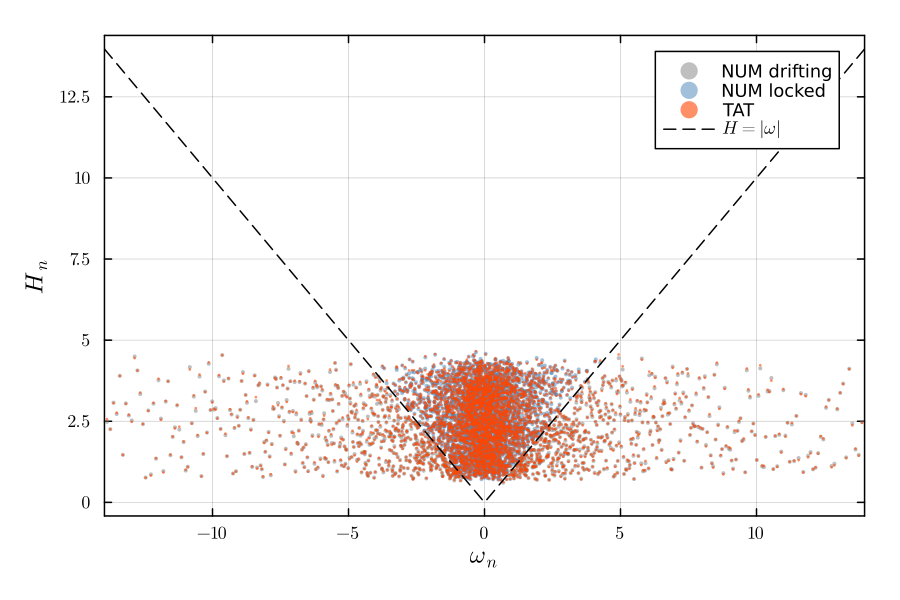}
        \caption{$K_2 = 1.5 K_2^c$}
        \label{fig:Hn_omega_k2_15}
    \end{subfigure}
    \caption{Scatter plots of the local mean field $H_n$ versus intrinsic frequency $\omega_n$ for a hypergraph with $N = 5000$, a uniform degree distribution in $[50, 250]$ ($\langle k \rangle = \langle q \rangle = 150$), and $K_3 = K_3^c$ at (a) subcritical dyadic coupling $K_2 = 0.8 K_2^c$ and (b) supercritical dyadic coupling $K_2 = 1.5 K_2^c$. Gray and blue circles denote drifting and locked oscillators, respectively, from direct numerical simulations of Eq.~\eqref{eq:ode_sys}, while orange dots represent the TAT prediction from Eqs.~\eqref{eq:h_tat_system}--\eqref{eq:r3_tat_system}. The black dashed lines mark the locking boundary $H_n = \vert{}\omega_n\vert{}$.}
    \label{fig:Hn_vs_omega}
\end{figure}

To empirically validate the assumption that $H_n$ and $\omega_n$ are statistically independent, Fig.~\ref{fig:Hn_vs_omega} shows scatter plots of the local order parameter $H_n$ against the natural frequency $\omega_n$ for a hypergraph of size $N = 5000$ with a uniform degree distribution on $[50, 250]$ ($\langle k \rangle = \langle q \rangle = 150$) and $K_3 = K_3^c$, comparing the subcritical regime ($K_2 = 0.8 K_2^c$, Fig.~\ref{fig:Hn_omega_k2_08}) and the supercritical regime ($K_2 = 1.5 K_2^c$, Fig.~\ref{fig:Hn_omega_k2_15}). Across both regimes, the horizontal band structure of the point clouds confirms that the distribution of $H_n$ shows no systematic dependence on $\omega_n$, directly supporting our factorization of their joint distribution.

Furthermore, comparing the two panels illustrates the distinct dynamical regimes on either side of the synchronization transition:
\begin{itemize}
    \item \textbf{Subcritical regime ($K_2 = 0.8 K_2^c$, Fig.~\ref{fig:Hn_omega_k2_08}):} Below the critical coupling strength, almost all oscillators lie outside the locking cone $H_n = \vert{}\omega_n\vert{}$ (black dashed lines) and are drifting (gray circles), with only a narrow sliver of oscillators near $\omega_n \approx 0$ transiently locked (blue circles). In this incoherent state, the TAT fixed-point solution correctly predicts $H_n \approx 0$ (orange dots), whereas the numerical simulation exhibits a small, nonzero baseline ($H_n \in [0.08, 0.25]$) driven by $\mathcal{O}(k_n^{-1/2})$ finite-size fluctuations from the drifting oscillators.
    \item \textbf{Supercritical regime ($K_2 = 1.5 K_2^c$, Fig.~\ref{fig:Hn_omega_k2_15}):} Above the onset of synchronization, the coherent mean field grows substantially ($H_n \in [0.7, 4.6]$), widening the locking cone $H_n \ge \vert{}\omega_n\vert{}$ so that a large fraction of oscillators become phase-locked (blue circles inside the cone) while high-frequency oscillators outside the cone continue to drift (gray circles). In this synchronized regime, coherent contributions dominate over finite-size fluctuations, and the TAT predictions (orange dots) closely match the numerical values of $H_n$ across both locked and drifting oscillators.
\end{itemize}

The third term of Eq.~(\ref{eq:each_term_R2}) is 
\begin{align}
\sum_{|\omega_m| > H_m} \boldsymbol{A}_{nm} \langle e^{i\theta_m}\rangle_t,
\end{align}
which corresponds to a sum over drifting oscillators. This term vanishes under the same independence and symmetry assumptions mentioned above, as shown in Appendix A of Ref.~\cite{Restrepo2005}.
\section{Derivation of Equation~(\ref{eq:r3_tat_system})} \label{sec:r3_tat_system} \

In this Appendix we provide a derivation for the expression for the triadic order parameter, $R_n^{(3)}$, in the TAT [Eq.~(\ref{eq:r3_tat_system})]. For $R_n^{(3)}(LD)$ and $R_n^{(3)}(DD)$ in Eqs.~(\ref{eq:R3_LD}) and (\ref{eq:R3_DD}), using the assumption that pair correlations can be neglected, we have 
\begin{align}
R_n^{(3)}(LD) &\approx \sum_{\substack{|\omega_m| \le H_m \\ |\omega_j| > H_j}}\boldsymbol{B}_{nmj}\langle e^{ 2i\theta_m}\rangle_t \langle e^{-i\theta_j}\rangle_t,\label{eq:R3_LD2}\\
R_n^{(3)}(DD) &\approx \sum_{\substack{|\omega_m| > H_m \\ |\omega_j| > H_j}}\boldsymbol{B}_{nmj}\langle e^{ 2i\theta_m}\rangle_t \langle e^{-i\theta_j}\rangle_t \label{eq:R3_DD2}
\end{align}
For the locked oscillators we use $\langle e^{i\theta}\rangle_t = e^{i\theta^*}$, while for the drifting oscillators we average over their stationary distribution, which has density \cite{Strogatz2000}
$$
\rho(\theta,\omega, H) = \frac{\sqrt{\omega^2-H^2}}{2\pi |\omega - H\sin(\theta)|}.
$$ 
We have, then,
\begin{align}
R_n^{(3)}(LD) &\approx \sum_{\substack{|\omega_m| \le H_m \\ |\omega_j| > H_j}}\boldsymbol{B}_{nmj}e^{ 2i\theta_m^*} \int_{0}^{2\pi} e^{-i\theta}\frac{\sqrt{\omega_j^2-H^2_j}}{2\pi |\omega_j - H_j\sin(\theta)|}d\theta,
\end{align}
As in the previous Appendix, we average over the frequency distribution, obtaining
\begin{align}
R_n^{(3)}(LD) &\approx \sum_{|\omega_m| \le H_m,j}\boldsymbol{B}_{nmj}e^{ 2i\theta_m^*} \int_{-H_j}^{H_j}\int_{0}^{2\pi} e^{-i\theta}\frac{g(\omega)\sqrt{\omega^2-H^2_j}}{2\pi |\omega - H_j\sin(\theta)|}d\theta d\omega = 0,
\end{align}
by the symmetry of $g(\omega)$ \cite{Strogatz2000}.
Similarly, $R_n^{(3)}(DD) \approx 0$. This eliminates the locked-drifting ($LD$) and drifting-drifting ($DD$) contributions, leaving only the locked-locked ($LL$) and drifting-locked ($DL$) terms.
Next we evaluate the term $R_n^{(3)}(LL)$:
\begin{align}
R_n^{(3)}(LL) &\approx \sum_{\substack{|\omega_m| \le H_m \\ |\omega_j| \le H_j}}\boldsymbol{B}_{nmj}\langle e^{ 2i\theta_m}\rangle_t \langle e^{-i\theta_j}\rangle_t,\\
& = \sum_{\substack{|\omega_m| \le H_m \\ |\omega_j| \le H_j}}\boldsymbol{B}_{nmj}e^{ 2i\theta^*_m}  e^{-i\theta_j^*} \\
&= \sum_{\substack{|\omega_m| \le H_m \\ |\omega_j| \le H_j}}\boldsymbol{B}_{nmj}\left[ 1- 2 \sin^2(\theta_m^*) + i 2\sin(\theta_m^*)\right]\left[\cos(\theta_j^*) + i \sin(\theta_j^*)\right].
\end{align}
Using $\sin(\theta^*) = \omega/H$, averaging again over the frequency distribution to estimate the terms with $\sin(\theta^*_m)$ and $\sin(\theta^*_j)$, and using the symmetry of $g(\omega)$, the only term that doesn't vanish is
\begin{align}
R_n^{(3)}(LL) &\approx  \sum_{\substack{|\omega_m| \le H_m \\ |\omega_j| \le H_j}}\boldsymbol{B}_{nmj}\left[ 1 - 2 \sin^2(\theta_m^*) \right]\cos(\theta_j^*)\nonumber\\
&= \sum_{\substack{|\omega_m| \le H_m \\ |\omega_j| \le H_j}} \boldsymbol{B}_{nmj} \left[1-2\left(\frac{\omega_m}{H_m}\right)^2\right]\sqrt{1-\left(\frac{\omega_j}{H_j}\right)^2}. 
\end{align}
Finally, we evaluate the term $R_n^{(3)}(DL)$, averaging the drifting oscillators over the stationary distribution $\rho(\theta,\omega,H)$:
\begin{align}
R_n^{(3)}(DL) &\approx \sum_{\substack{|\omega_m| > H_m \\ |\omega_j| \le H_j}}\boldsymbol{B}_{nmj}\langle e^{ 2i\theta_m}\rangle_t \langle e^{-i\theta_j}\rangle_t,\\
& \approx \sum_{\substack{|\omega_m| > H_m \\ |\omega_j| \le H_j}}\boldsymbol{B}_{nmj}\left[\int_{0}^{2 \pi}  e^{ 2i\theta}\rho(\theta,\omega_m,H_m)d\theta\right] e^{-i\theta_j^*}\\
& = \sum_{\substack{|\omega_m| > H_m \\ |\omega_j| \le H_j}}\boldsymbol{B}_{nmj} \left[\frac{2|\omega_m|}{H_m^2}\left(\sqrt{\omega_m^2-H_m^2}-\omega_m\right)+1 \right]\left[\cos(\theta_j^*) + i \sin(\theta_j^*)\right]\\
& = \sum_{\substack{|\omega_m| > H_m \\ |\omega_j| \le H_j}}\boldsymbol{B}_{nmj} \left[\frac{2|\omega_m|}{H_m^2}\left(\sqrt{\omega_m^2-H_m^2}-\omega_m\right)+1 \right]\sqrt{1-\left(\frac{\omega_j}{H_j} \right)^2},
\end{align}
where, again, the imaginary term vanishes by symmetry of the frequency distribution.

Combining these derivations, we assemble the final expression for $R_n^{(3)}$ in Eq.~(\ref{eq:r3_tat_system}):
\begin{align}
    R_n^{(3)} &= R_n^{(3)}(LL) + R_n^{(3)}(DL) \nonumber \\
    &= \sum_{\substack{|\omega_m| \le H_m \\ |\omega_j| \le H_j}} \boldsymbol{B}_{nmj} \left[1-2\left(\frac{\omega_m}{H_m}\right)^2\right]\sqrt{1-\left(\frac{\omega_j}{H_j}\right)^2} \nonumber \\
    &\quad + \sum_{\substack{|\omega_m| > H_m \\ |\omega_j| \le H_j}} \boldsymbol{B}_{nmj} \left[\frac{2|\omega_m|}{H_m^2}\left(\sqrt{\omega_m^2-H_m^2}-\omega_m\right)+1 \right]\sqrt{1-\left(\frac{\omega_j}{H_j} \right)^2}.
\end{align}

\section{FDA derivation} \label{appx:FDA_integral}

In this Appendix we clarify the procedure for obtaining Equation~(\ref{eq:r3_fda}) from Eq.~(\ref{eq:r3_tat_system}). Following Eq.~(\ref{intapp2}), we approximate the sums as integrals,
\begin{align}
      R^{(3)}_n (LL)  &\approx \sum_{m,j} \boldsymbol{B}_{nmj} I_1(H_m) I_2(H_j),
\end{align}
where
\begin{align}
    I_1(H_m) &=  \int^{H_m}_{-H_m} \ g(\omega) \ \left[ 1- 2\left( \frac{\omega}{H_m} \right)^2 \right]d\omega = \frac{2}{\pi H_m^2} \left[(H_m^2+2)\arctan(H_m)-2H_m\right] \\
    I_2(H_j) &= \int^{H_j}_{-H_j} \ g(\omega) \ \sqrt{1-\left(\frac{\omega}{H_j} \right)^2}d\omega = \frac{\sqrt{H_j^2+1}-1}{H_j} \label{i1}
\end{align}
Therefore, 
\begin{align}
    R^{(3)}_n(LL) &\approx  \sum_{m=1}^N\sum_{j=1}^N \boldsymbol{B}_{nmj} \left( \frac{2}{\pi H_m^2} \left[(H_m^2+2)\arctan(H_m)-2H_m\right] \right) \left( \frac{\sqrt{H_j^2+1}-1}{H_j}  \right) \label{rll}
\end{align}
Now we focus on the term $R^{(3)}_n (DL)$:
\begin{align}
 R^{(3)}_n(DL) &= \sum_{\substack{|\omega_m| > H_m \\ |\omega_j| \le H_j}} \boldsymbol{B}_{nmj} \langle e^{2i\theta_m}e^{-i\theta_j} \rangle_t \\
 & \approx \sum_{\substack{|\omega_m| > H_m \\ |\omega_j| \le H_j}} \boldsymbol{B}_{nmj} \langle e^{2i\theta_m}\rangle_t \langle e^{-i\theta_j} \rangle_t\\
    &\approx   \sum_{\substack{|\omega_m| > H_m \\ |\omega_j| \le H_j}} \boldsymbol{B}_{nmj} \langle e^{2i\theta_m}\rangle_t \left[ \cos(\theta^*_j) + i \sin(\theta^*_j)\right].
\end{align}
The term with $\sin(\theta_j^*)$ vanishes, again, and we get
\begin{align}
    R^{(3)}_n(DL)  &\approx   \sum_{m,j} \boldsymbol{B}_{nmj} I_3(H_m) I_2(H_j),\label{rdli1i5}
\end{align}
where the integral $I_3$ is again done using the stationary distribution of drifting oscillators
\begin{align}    
    I_3(H)  &=\int^\infty_H  \int_0^{2\pi} \frac{e^{-2i\theta}}{\pi(1+\omega^2)}  \rho(\theta,\omega,H) d\theta d\omega+ \int^{-H}_{-\infty} \int_0^{2\pi} \frac{e^{-2i\theta}}{\pi(1+\omega^2)}  \rho(\theta,\omega,H) d\theta d\omega\\
    & = 2\left[ \frac{2H- \pi\sqrt{H^2+1}+(H^2+2)(\pi/2+\arctan(H))}{\pi H^2} \right].\label{i5}
\end{align}
Now we can replace $I_1$ from Eq.~(\ref{i1}) and $I_3$ from Eq.~(\ref{i5}) in Eq.~(\ref{rdli1i5}) to obtain $R^{(3)}_n(DL)$:
\begin{align}
    R^{(3)}_n(DL) &=  \sum_{m=1}^N\sum_{j=1}^N \boldsymbol{B}_{nmj}  2\left[ \frac{2H_m- \pi\sqrt{H_m^2+1}+(H_m^2+2)(\pi/2-\arctan(H_m))}{\pi H_m^2} \right]   \frac{\sqrt{H_j^2+1}-1}{H_j}\label{rdl}
\end{align}
Substituting Eqs.~(\ref{rll}), (\ref{rdl}) in $R_n^{(3)} \approx R_n^{(3)}(LL) + R_n^{(3)}(DL)$ and simplifying, we find $R_n^{(3)}$:
\begin{align}
    R^{(3)}_n &=  \sum_{m=1}^N\sum_{j=1}^N\boldsymbol{B}_{nmj}  \frac{\sqrt{H_j^2+1}-1}{H_j} \cdot \Bigg[ 2\frac{2H_m- \pi\sqrt{H_m^2+1}+(H_m^2+2)(\pi/2-\arctan(H_m))}{\pi H_m^2} \nonumber \\
    &+ \frac{2}{\pi H_m^2} ((H_m^2+2)\arctan(H_m)-2H_m) \Bigg] \\
    &=  \sum_{m=1}^N\sum_{j=1}^N\boldsymbol{B}_{nmj} \left( \frac{\sqrt{H_m^2+1}-1}{H_m} \right)^2\frac{\sqrt{H_j^2+1}-1}{H_j},
\end{align}
which is Equation~(\ref{eq:r3_fda}) in the main text.

\section{Rewiring Algorithm}
\label{appx:rewire_alg} \

In this Appendix we describe the rewiring algorithm that we implement in Sec.~\ref{rewiring} to modify the onset of bistability. According to Equation~(\ref{eq:K3c_eq}), the critical coupling strength $K_3^c$ for the onset of bistability is influenced by correlations between the left and right Perron-Frobenius eigenvectors of the adjacency matrix $\boldsymbol{A}$ and the triadic adjacency tensor $\boldsymbol{B}$. In particular, for a fixed adjacency matrix $\boldsymbol{A}$, $K_3^c$ decreases (increases) if the sum 
\begin{align}
\mathcal{D} = \sum_{n,m,j} V_n \boldsymbol{B}_{nmj} U_m^2 U_j
\end{align}
increases(decreases).
\begin{figure}[t]
    \centering  
    \begin{subfigure}{0.4\textwidth}
        \centering
        \includegraphics[width=\linewidth]{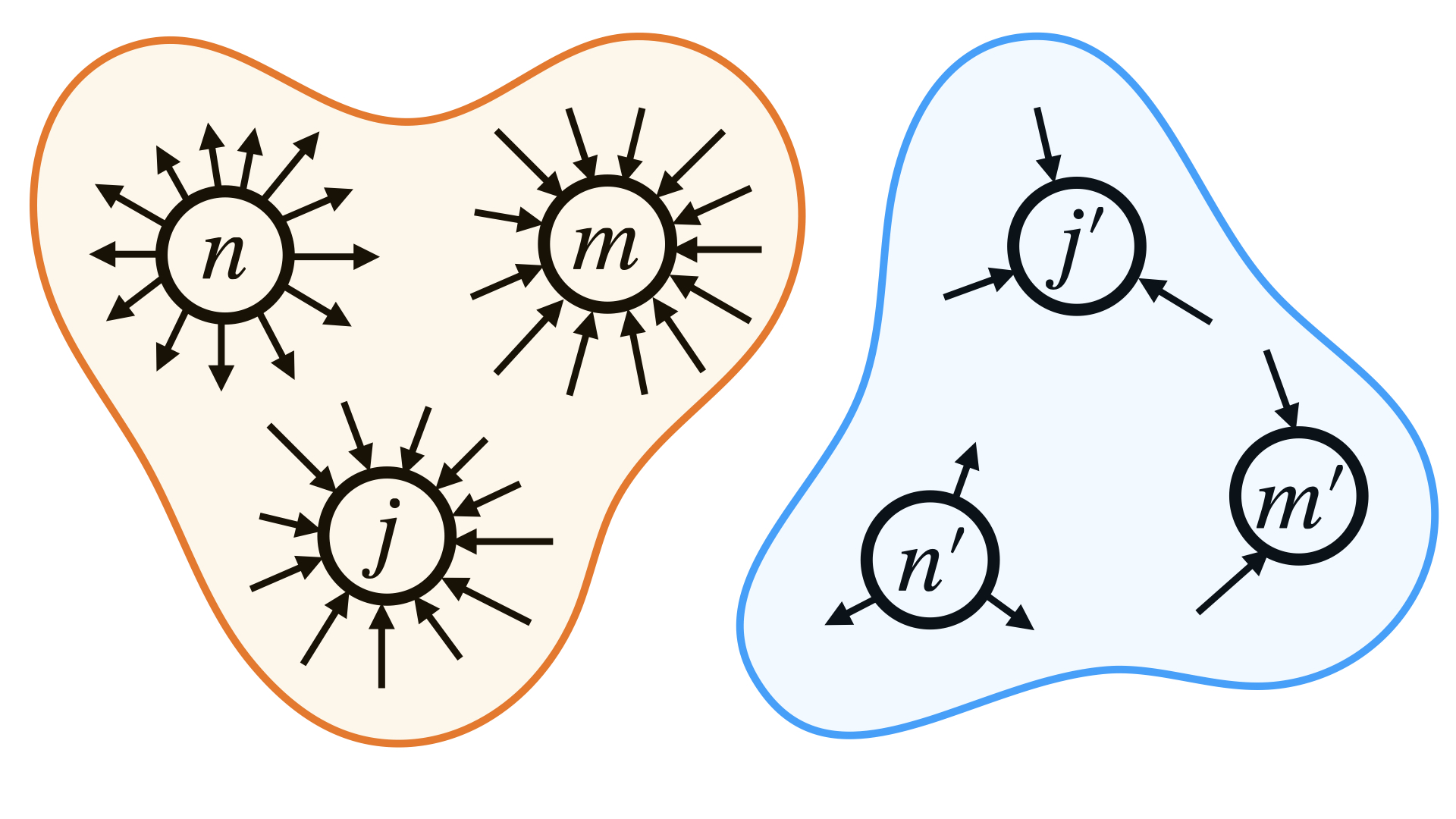}
        \caption{}
        \label{fig:rewirea}
    \end{subfigure}
    \hfill
    \begin{subfigure}{0.4\textwidth}
        \centering
        \includegraphics[width=\linewidth]{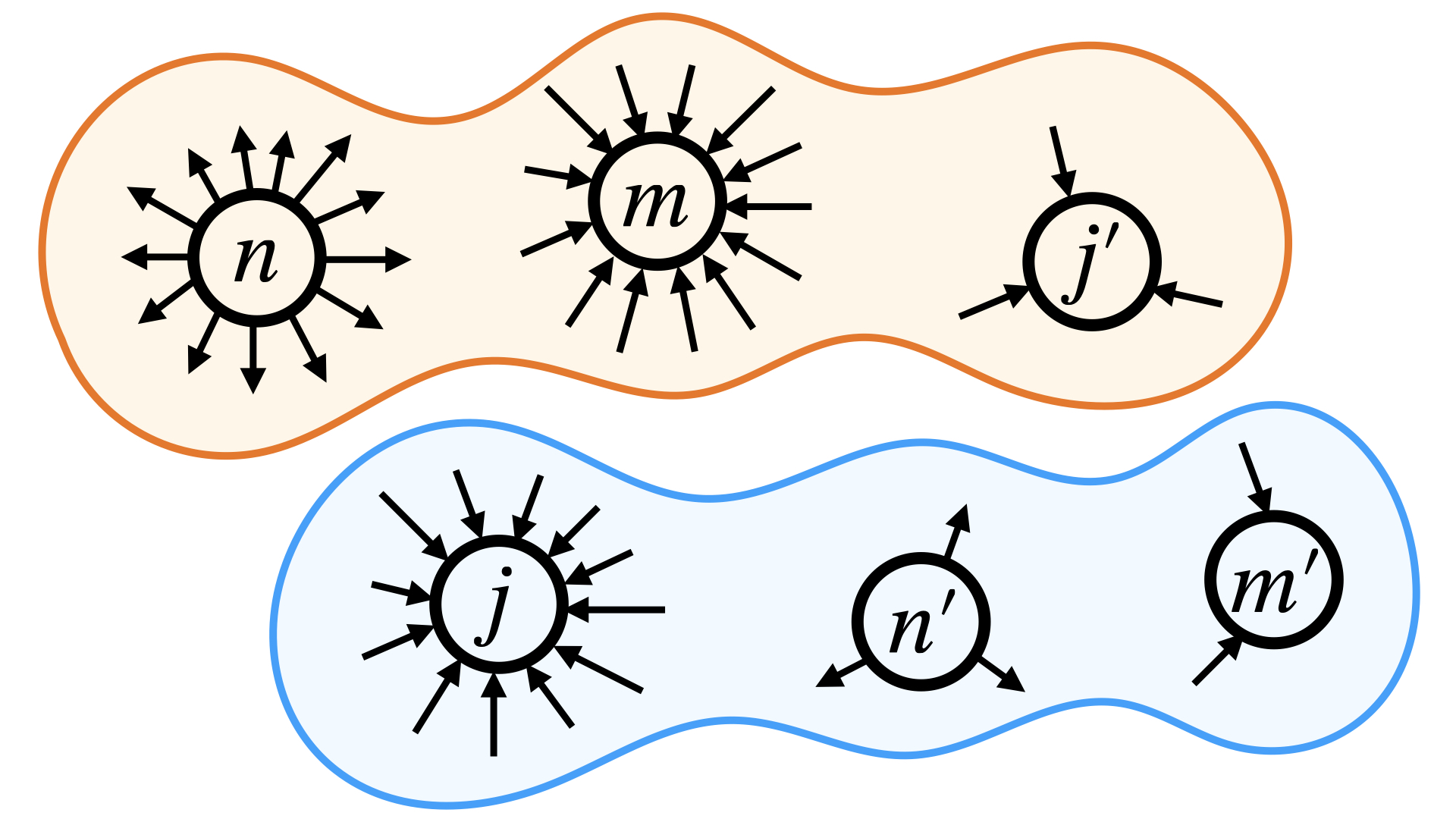}
        \caption{}
        \label{fig:rewireb}
    \end{subfigure}

    \caption{(a) Chosen hyperedges before rewiring. (b) Hyperedges with nodes $j$ and $j'$ swapped, to be considered for rewiring.}
    \label{fig:rewiring}
\end{figure}

To investigate how structural modifications affect $K_3^c$ we implement a targeted network rewiring procedure. The objective is to shift the subcritical point by maximizing or minimizing the interaction term $\mathcal{D}$ through a hypergraph rewiring process. The procedure, which is motivated by Newman's rewiring algorithm \cite{newman2003mixing}, and which has also recently been applied to hypergraphs~\cite{Landry2022}, is executed as follows:
\begin{enumerate}
    \item \textbf{Initialization:} Construct an initial hypergraph containing both dyadic and triadic interactions, encoded by dyadic and triadic adjacency tensors $ \boldsymbol{A}$ and  $\boldsymbol{B}$. We consider only unweighted hypergraphs for which $\boldsymbol{B}_{nmj} \in \{0,1\}$.
    \item \textbf{Edge Selection:} Randomly and uniformly select two nodes, say $n$ and $n'$, and identify their existing triadic connections. Select one triadic connection for each, say $\{n,m,j\}$ and $\{n',m',j'\}$ [see Figure~\ref{fig:rewiring}(a)]. If these two sets are not disjoint, repeat this step. 
    \item \textbf{Stochastic Swap:} Select one node from each of the two sets uniformly at random, say $j$ and $j'$, and construct a new adjacency tensor $\boldsymbol{B} \to \boldsymbol{\tilde B}$ by swapping these nodes as
    \begin{align}
      \{n,m,j\} & \to \{n,m,j'\}\\
      \{n',m',j'\} & \to \{n',m',j\}.
    \end{align}
    \item \textbf{Local Evaluation:} Calculate the original and updated local denominator terms:
    \begin{align}
        \mathcal{D} &= \sum_{n,m,j} V_n \boldsymbol{B}_{nmj} U_m^2 U_j,\\
        \mathcal{\tilde D} &= \sum_{n,m,j} V_n \boldsymbol{\tilde B}_{nmj} U_m^2 U_j
    \end{align}
    \item \textbf{Acceptance Criterion:} 
    \begin{itemize}
        \item To \textit{lower} the critical point $K_3^c$ relative to the original network, accept the swap only if $\mathcal{\tilde D} > \mathcal{D}$. 
        \item To \textit{raise} the critical point $K_3^c$, accept the swap only if $\mathcal{\tilde D} < \mathcal{D}$.
    \end{itemize}
    If the swap is accepted, let $\boldsymbol{B} = \boldsymbol{\tilde B}$; otherwise, do nothing. 
    \item \textbf{Iteration:} Repeat the process until all nodes have been considered or the system reaches the desired threshold shift.
\end{enumerate}

Note that steps 4 and 5 can be optimized by only comparing the relevant parts of the sums, i.e., it is enough to compare $d$ and $\tilde d$, where
    \begin{align}
        d &= V_n U_m^2 U_j + V_{n'} U_{m'}^2 U_{j'},\\
        \tilde d &= V_n U_m^2 U_{j'} + V_{n'} U_{m'}^2 U_{j}.
    \end{align}
While one can improve the rewiring algorithm by introducing an acceptance probability to avoid getting stuck in local maxima or minima, we didn't implement that feature.

\end{document}